\documentclass[11pt]{article} 
\usepackage[top=20mm,bottom=20mm, left=20mm, right=20mm]{geometry}

\usepackage{authblk}
\usepackage{enumerate}
\usepackage{amsmath, cite, mathrsfs}
\usepackage{graphicx, color, hyperref}
\usepackage{amssymb, amsthm, makecell, enumitem}
\usepackage{stmaryrd}
\usepackage{subcaption}

\newcounter{dummy}

\makeatletter
\newcommand{\mylabel}[2]{#2\refstepcounter{dummy}\def\@currentlabel{#2}\label{#1}}
\makeatother

 \hypersetup{
     colorlinks=true,       % false: boxed links; true: colored links
     linkcolor=red,          % color of internal links (change box color with linkbordercolor)
     citecolor=blue,        % color of links to bibliography
     filecolor=magenta,      % color of file links
     urlcolor=cyan           % color of external links
 }
 
\theoremstyle{plain}
\newtheorem{theorem}{Theorem}[section]
\newtheorem{lemma}[theorem]{Lemma}

\newtheorem{id}{\bf Identity}[section]

\theoremstyle{remark}
\newtheorem{remark}{Remark}[section]

\def\dl{\operatorname{dl}}
\def\da{\operatorname{da}}

\def\inf{\operatorname{inf}}

\def\supp{\operatorname{supp}}
\def\singsupp{\operatorname{singsupp}}

\def\div{\operatorname{div}}
\def\Div{\operatorname{Div}}

\def\curl{\operatorname{curl}}
\def\Curl{\operatorname{Curl}}

\def\tr{\operatorname{tr}}
\def\Lin{\operatorname{Lin}}
\def\Sym{\operatorname{Sym}}
\def\Skw{\operatorname{Skw}}
\def\Det{\operatorname{Det}}
\def\det{\operatorname{det}}
\def\deg{\operatorname{deg}}

\title{Singular Points and Singular Curves in von K{\'a}rm{\'a}n Elastic Surfaces}
\author{Animesh Pandey and Anurag Gupta\thanks{ag@iitk.ac.in}}
\date{Department of Mechanical Engineering, Indian Institute of Technology Kanpur, 208016, India\\[2ex]%
    \today
}

\begin{document}
\maketitle

\begin{abstract}
Mechanical fields over thin elastic surfaces can develop singularities at isolated points and curves in response to constrained deformations (e.g., crumpling and folding of paper), singular body forces and couples, distributions of isolated defects (e.g., dislocations and disclinations), and singular metric anomaly fields (e.g., growth and thermal strains). With such concerns as our motivation, we model thin elastic surfaces as von K{\'a}rm{\'a}n plates and generalize the classical von K{\'a}rm{\'a}n equations, which are restricted to smooth fields, to fields which are piecewise smooth, and can possibly concentrate at singular curves, in addition to being singular at isolated points. The inhomogeneous sources to the von K{\'a}rm{\'a}n equations, given in terms of plastic strains, defect induced incompatibility, and body forces, are likewise allowed to be singular at isolated points and curves in the domain. The generalized framework is used to discuss the singular nature of deformation and stress arising due to conical deformations, folds, and folds terminating at a singular point.
\end{abstract}


\section{Introduction}
Thin elastic sheets, such as those made of paper, crumple by developing kinks, creases, folds, and singular stress fields in response to constrained deformations, e.g., while squeezing them inside a hollow sphere or crushing them in our fist~\cite{ben1997crumpled, lobkovsky1997properties, cerda1998conical, witten2007, efrati2015confined}. The singularities persist even after the constraints are released indicating that the incurred deformations are plastic in nature. Similarly, an isolated defect, such as a disclination or a dislocation, in an elastic sheet (e.g., two-dimensional (2D) crystalline and amorphous materials) yields a deformed shape with a cone like singularity and a singular stress both at the defect location~\cite{SeungNelson88, muller08, olbermann2017, PandeyPRE}. In this case the singularity appears due to an internal source of strain incompatibility (in the form of a defect) even in the absence of any external constraints and forces. The singular deformations and stress fields can also appear in response to concentrations of external forces and force couples, and singular sources of metric anomalies (e.g., thermal and growth strains). 
In the present work, we model elastic sheets as von K{\'a}rm{\'a}n plates and use the theory of distributions to develop a general theoretical framework within which we can discuss a large variety of problems where singularities emerge at isolated points and curves in the plate domain. The existing von K{\'a}rm{\'a}n equations, which assume the fields to be smooth, are extended by including equations which should be satisfied at singular points and singular curves. The sources of inhomogeneity, given in terms of plastic strains, defect densities, and body force fields, are accordingly allowed to be singular. We also allow for several fields to develop concentrations over the singular subdomain. To the best of our knowledge, any such generalization of the von K{\'a}rm{\'a}n framework has not been attempted previously in the literature.

In order to elaborate on the contributions of this work, we begin by stating the inhomogeneous von K{\'a}rm{\'a}n equations for smooth fields with inhomogeneities given in terms of smooth plastic strain fields, an incompatibility field (which can be obtained in terms of smooth defect densities), and a smooth body force field. A brief derivation of these equations, following our recent work~\cite{manish1, manish2}, has been given in Section~\ref{inhomovksmooth}. Consider a plate domain $\Omega$ over which all the considered fields are smooth. The two inhomogeneous von K{\'a}rm{\'a}n equations for transverse displacement $w$ and stress function $\phi$ are given as
\begin{subequations}
\label{introvk}
\begin{align}
\frac{1}{E}\Delta^2 \phi +\frac{1}{2} [w,w]= - \curl \curl \boldsymbol{e}^p = -\eta_1 + \det({\boldsymbol\lambda}^p)  ~ \text{in}~\Omega~\text{and} \label{intro1}
\\
D\Delta^2 w - [\phi,w]=f + D \big( (1-\nu)\div \div {\boldsymbol{\lambda}^p} +\nu \Delta \tr({\boldsymbol{\lambda}^p})\big)  ~ \text{in}~\Omega,\label{intro2}
\end{align}
\end{subequations}
where $E$, $D$, and $\nu$ are material parameters, $\boldsymbol{e}^p$ and  ${\boldsymbol\lambda}^p$ are plastic stretching and bending strains, respectively, $\eta_1$ is the strain incompatibility field, and $f$ is the transverse body force field. The Monge-Amp{\`e}re bracket $[\cdot,\cdot]$ is defined at the beginning of Section~\ref{mabrack}, while other notational details are given in Section~\ref{notation}.
Our aim is to generalize the inhomogeneous von K{\'a}rm{\'a}n equations \eqref{intro1} and  \eqref{intro2} to non-smooth fields. In particular, we allow the fields to be singular at a point $O \in \Omega$, piecewise smooth across a curve $S \subset \Omega$, and, in some cases, to concentrate along the singular curve $S$. The curve $S$ is assumed to be regular, except at point $O$ (i.e., in case $O\in S$). 

The first step in the proposed generalization is to consider the strain-displacement and the strain compatibility relations in the sense of distributions, see Section~\ref{weakcomp}. The difficulty in the former is due to the non-linear appearance of the transverse displacement in the stretching strain field. We make specific assumptions on the singular nature of $w$ to posit the strain-displacement relation in a reasonable form. Essentially, we assume $w$ to remain bounded at $O$. The derivatives of $w$ can become unbounded at $O$. Similarly, in order to state the compatibility equations as a distributional relation, we make suitable assumptions on the bending strain such that the distributional determinant is well defined. We note that compatibility conditions for non-smooth strain fields (in the context of von K{\'a}rm{\'a}n plates) have been discussed earlier ~\cite{ciarlet2013nonlinearDonati,ciarlet2013nonlinear}, assuming stretching and bending strain fields to be square integrable,  in terms of the distributional derivatives (not the strong forms, as derived in this paper). The assumption of square integrability allows for strain fields to be discontinuous across $S$, however it precludes the possibility of a concentration of the bending strain on the singular interface. The concentration of bending strain is  necessary in modelling sharp folds where the normal to the surface jumps across the interface (such folds appear commonly in a crumpled paper). The explicit form of pointwise (in $\Omega - \{S \cup \{O\}\}$, on $S-\{O\}$, and at $O$)  strain-displacement relations and strain compatibility relations are given in Equations~\eqref{DefinitionofStrains} and \eqref{CurlLambdaCompatibility}-\eqref{GaussiamCurvatureCompatibiltiy}, respectively. 

Subsequently, we introduce plastic strain fields as distributions with an allowance for interfacial concentrations in plastic bending strain, see Section~\ref{sources}. The generalized incompatibility relations for the plastic strains are used to argue that either plastic stretching and bending strains or an incompatibility field and the plastic bending strain field can be prescribed as sources of inhomogeneity arising due to strain incompatibility. The generalized incompatibility field is related to bulk, interfacial, and point concentrations of defect densities, see Equations~\eqref{N1Restriction}. The next step is to posit the equilibrium equations in a distributional form, see Section~\ref{geneqb}. Both in-plane stress and moment are expressed in terms of a bulk density and an interfacial concentration. The main challenge in posing the distributional equilibrium equations is to have a well defined (distributional) inner product between stress and bending strain fields, a non-linear term which appears in the balance of transverse forces over $\Omega$. Towards this end, specific regularity assumptions are made on $w$ and $\phi$ (similar to those made while stating the compatibility equations) so that the distributional inner product is well defined. We assume the external transverse forces to be given in terms of a bulk density, an interfacial concentration on $S$, and a point supported concentration at $O$. The transverse force couples (and higher order multipoles) are also incorporated in the distributional framework. The pointwise equilibrium conditions, in terms of various stress and moment fields, over $\Omega$ (away from $\{S \cup \{O\}\}$), on $S-\{O\}$, and at $O$, are given in Equations~\eqref{EquilibriumEquationsDiscontinuousStressFields} and \eqref{MomentBalanceInterface}. Under certain regularity assumptions the transverse force balance at $O$ can be written as a loop integral around $O$.

The desired (distributional) von K{\'a}rm{\'a}n equations are obtained by introducing further regularity assumptions and constitutive restrictions, see Section~\ref{genvkequations}. We assume that both stress and moment fields do not concentrate on $S$. The elastic strain fields are also assumed to not concentrate on $S$. Therefore the concentrations in total and plastic bending strain on $S$ are identical. Moreover, any concentration in plastic stretching strain is not permitted. The bulk stress and moment are related to bulk elastic strain fields through a linear, isotropic, and materially uniform constitutive relationship. The von K{\'a}rm{\'a}n equations are derived using the compatibility relations, the prescription of inhomogeneity sources, equilibrium equations, and constitutive assumptions. The equations are distinguished based on whether the plastic bending strain is prescribed with plastic stretching strain or with an incompatibility field. For the former case, the distributional von K{\'a}rm{\'a}n equations are given in \eqref{vk1dist1} and \eqref{vk2dist1} or, equivalently, as local pointwise equations in \eqref{plastbendstrconc}, \eqref{vk1loc}, and \eqref{VonKArmanEquilibriumStrongForms}. For the latter case, the distributional equations are given in \eqref{vk2dist1} and \eqref{vk1pdist1} or, equivalently, as local pointwise equations in \eqref{vk1ploc} and \eqref{VonKArmanEquilibriumStrongForms1}. The local equations in the bulk, away from $\{S \cup \{O\}\}$, are of the same form as those given for smooth fields (as in \eqref{introvk}). The local equations at the singular interface (i.e., on $S-\{O\}$) and the singular point $O$ are novel and form the central contribution of this paper. 

In Section~\ref{applications} we revisit some singular problems within the context of our developed framework. We discuss four types of problems, all with the assumption of elastic inextensibility (i.e., vanishing of elastic stretching strains over $\Omega$). First, we consider a plate domain with a singularity only at $O$, see Section~\ref{conical}. Such a scenario can emerge if there is an isolated defect at $O$~\cite{SeungNelson88, PandeyPRE}, or if there is a point supported transverse force (or force couple) at $O$, or during constrained deformations~\cite{ben1997crumpled, cerda1998conical, witten2007}. Only in the first case, there will be a non-trivial Gaussian curvature at $O$. We establish that, for a compatible bending strain, with a singular support at $O$, and a Gaussian curvature, which necessarily vanishes outside $O$, the deformation is conical (i.e., of the form $w=rg(\theta)$). In another result we demonstrate that a conical deformation, with a logarithmic stress function (i.e., $\phi$ proportional to $\ln r$), satisfies equilibrium only in the absence of bulk and point supported external forces (or couples). Several issues associated with the singular solution to the disclination problem and the constrained deformation problem are discussed. Secondly, in Section~\ref{folds}, we allow the plate domain to develop a fold, as a concentration in bending strain, without any point $O$ of singularity.  The fold $S$ can either close onto itself or end at the boundary of the domain. We first reduce the general  von K{\'a}rm{\'a}n equations to the case at hand and then discuss solutions for a linear and a circular fold. Third, we consider a linear fold whose one end is on the boundary of $\Omega$ and the other end is at an internal point $O$ inside the domain~\cite{lechenault2015generic}, see Section~\ref{ridge}. Therefore we have a situation where we have a singular interface terminating at a singular point. We establish the validity of the conical deformation, with $g$ continuous but piecewise smooth across the fold, and a logarithmic stress function. In all our problems, we verify that the considered solutions satisfy the local equations in bulk, on the interface, and at the singular point. Finally, in Section~\ref{tetra}, we consider a problem where several straight folds terminate at an internal singular point $O$ in the domain, while having their other end point on the boundary of the domain. The strength of the point supported Gaussian curvature at $O$ is calculated in terms of the strength and orientation of the intersection folds.

The present paper builds upon the distributional framework that was developed to study singular solutions in linear elasticity theories in our recent work~\cite{pandey2020topological, pandey2022point}. One of these was concerned with deriving strain compatibility and incompatibility relations for piecewise smooth linearized strain fields, possibly with concentrations over the singular interfaces~\cite{pandey2020topological}, while the other dealt with equations of equilibrium and strain compatibility/incompatibility for fields with point singularities~\cite{pandey2022point}. In the latter, emphasis was placed in rigorously establishing the gap between stating the governing equations in terms of the bulk restriction of the fields (away from the singular points) and in distributional terms over the whole domain. The notion of degree of divergence was central in characterizing the gap between the two descriptions. The present work significantly deviates from these papers due to the inherently non-linear nature of the von K{\'a}rm{\'a}n equations. In order to pose the von K{\'a}rm{\'a}n equations in a distributional sense, we need to unambiguously establish the notions of a distributional determinant, a distributional inner product, and a distributional Monge-Amp{\`e}re bracket, while remembering that the multiplication of two arbitrary distributions is in general not defined, see Section~\ref{mabrack}. Towards this end, we have to specify stricter regularity assumptions (including those related to the degree of divergence) on the associated fields. Overall, the present paper demonstrates a successful application of a distributional methodology to a non-linear framework with singular fields. Most importantly, it allows us to obtain local pointwise equations that hold over the singular subdomain. Our work is in fact a clear departure from the classical elasticity solutions where the application of distribution theory has been mostly limited to the representation of singular fields in terms of Dirac delta and its gradients~\cite{podio2014elasticity}.

\section{Mathematical preliminaries}

\subsection{Notation} \label{notation}

Given any two vectors $\boldsymbol{u},\boldsymbol{v} \in \mathbb{R}^2$, let $\langle\boldsymbol{u},\boldsymbol{v}\rangle$ represent their inner product such that $\langle\boldsymbol{u},\boldsymbol{v}\rangle=u_i v_i$ (summation over repeated indices), where $u_i=\langle\boldsymbol{u},\boldsymbol{e}_i \rangle$. The pair $\{ \boldsymbol{e}_1, \boldsymbol{e}_2\}$ is a fixed orthonormal basis in $\mathbb{R}^2$. Let $\boldsymbol{e}_3$ be a unit vector such that, for any $\boldsymbol{v}\in\mathbb{R}^2$, $\boldsymbol{e}_3\times\boldsymbol{v}=-\boldsymbol{v}\times \boldsymbol{e}_3=-v_2\boldsymbol{e}_1+v_1\boldsymbol{e}_2$. The space of linear mappings from $\mathbb{R}^2$ to itself (second order tensors) is denoted by $\Lin$, the space of symmetric tensors by $\Sym$, and the space of skew symmetric tensors by $\Skw$. Given $\boldsymbol{u},\boldsymbol{v} \in \mathbb{R}^2$,  the dyadic product $\boldsymbol{u} \otimes \boldsymbol{v} \in \Lin,$ is defined such that $(\boldsymbol{u} \otimes \boldsymbol{v}) \boldsymbol{w}=\langle \boldsymbol{v}, \boldsymbol{w} \rangle \boldsymbol{u}$ for all $\boldsymbol{w}\in \mathbb{R}^2.$  Given two tensors $\boldsymbol{a},\boldsymbol{b}\in \Lin$ their inner product $\langle \boldsymbol{a},\boldsymbol{b}\rangle$ is such that, for $\boldsymbol{u}, \boldsymbol{v}, \boldsymbol{w}, \boldsymbol{q} \in \mathbb{R}^2$, we have $\langle \boldsymbol{u} \otimes \boldsymbol{w},\boldsymbol{v} \otimes \boldsymbol{q} \rangle=\langle \boldsymbol{u}, \boldsymbol{v} \rangle \langle \boldsymbol{w}, \boldsymbol{q} \rangle$. The identity tensor is written as $\boldsymbol{I}\in \Lin$. For any tensor $\boldsymbol{a}\in \Lin$, $\tr(\boldsymbol{a})=\langle \boldsymbol{a},\boldsymbol{I} \rangle$ represents the trace of $\boldsymbol{a}$ while $\det(\boldsymbol{a})$ represents the determinant.  We define a linear operator $\mathbb{A}:\Lin \to \Lin$ such that, for $\boldsymbol{v}, \boldsymbol{w} \in \mathbb{R}^2$, $\mathbb{A} (\boldsymbol{v}\otimes \boldsymbol{w})=(\boldsymbol{e}_3\times \boldsymbol{v})\otimes(\boldsymbol{e}_3\times \boldsymbol{w})$.

Given two sets $A$ and $B$, $A\subset B$ indicates that $A$ is a subset of $B$ and $A-B$ represents the difference of $A$ and $B$. We use $\Omega \subset \mathbb{R}^2$ to represent an open, connected set. Let $O \in \Omega$ be a point in the interior of $\Omega$. The set $\Omega-O$ is identical to $\Omega-\{O\}$, i.e., the difference of $\Omega$ and the singleton set $\{O\}$. We use $B_\epsilon$ to represent an open ball of radius $\epsilon>0$ centered at $O$ with $\epsilon$ such that $B_\epsilon$ (and its closure) is contained inside $\Omega.$ Let $\Omega_\epsilon=\Omega-B_\epsilon$. We use $S\subset \Omega$ to represent a smooth oriented curve in $\Omega$ with unit tangent vector $\boldsymbol{t}$ and unit normal vector $\boldsymbol{\nu}$. Given a differentiable field $f$ on $S$, ${df}/{ds}$ is the derivative of $f$ along the curve where $s$ is the arc length parameter. The curvature of $S$ is denoted by $k$, i.e., $k=\langle {d\boldsymbol{t}}/{ds}, \boldsymbol{\nu} \rangle$. We will use $S$ and $O$ to represent a curve and a point of singularity, respectively, in $\Omega$. The point $O$ may or may not lie on $S$. If $O\in S,$ then $O$ can be point of discontinuity for fields (including $\boldsymbol{t}$) over $S$. We assume that $\partial S - \partial \Omega \subset \{O\}$, i.e., the curve $S$ cannot end inside the domain $\Omega$ (except at point $O$). The area measure on $\Omega$ is represented by $\da$ and the length measure on curves in $\Omega$ is represented by $\dl$. Given any open set $\omega\subset \Omega,$ $C^\infty (\omega),$ $C^\infty (\omega,\mathbb{R}^2)$, and $C^\infty (\omega,\Lin)$ are the spaces of smooth scalar, vector, and tensor valued fields on $\omega$, respectively. Given a differentiable function $f$ on $\Omega$, $\partial f/\partial x_i $ is the partial derivative of $f$ with respect to the $i$-th component of the position vector $\boldsymbol{x} \in \Omega$. The gradient of $f$ is given by $\nabla f=(\partial f/\partial x_i)  \boldsymbol{e}_i$. The divergence of a differentiable vector field $\boldsymbol{f}\in C^1(\Omega,\mathbb{R}^2)$ is a continuous scalar valued field given by $\div \boldsymbol{f}=\partial f_1/\partial x_1 +\partial f_2/\partial x_2$. The curl of a differentiable vector field $\boldsymbol{f}\in C^1(\Omega,\mathbb{R}^2)$ is a continuous scalar valued field given by $\curl \boldsymbol{f}=\partial f_2/\partial x_1 - \partial f_1/\partial x_2$. The Laplacian of a twice differentiable scalar field, $f \in C^2(\Omega)$ is a continuous scalar valued field given by $\Delta f=\partial^2 f/\partial x_1 ^2 + \partial^2 f/\partial x_2 ^2$. We will occasionally use polar coordinates $(r,\theta)$, with $r = (x_1^2 + x_2^2)^{1/2}$ and $\theta = \tan^{-1}(x_2/x_1)$, $O$ as the centre, and polar basis $\{ \boldsymbol{e}_r, \boldsymbol{e}_\theta\}$.

\subsection{Distributions}

For any open set $\Omega\subset \mathbb{R}^2,$ let $\mathcal{D}(\Omega)$ be the space of compactly supported smooth functions on $\Omega$. The space of distributions $\mathcal{D}'(\Omega)$ is defined as the dual space of $\mathcal{D}(\Omega)$~\cite{friedlander1998introduction}. Similarly, $\mathcal{D}'(\Omega,\mathbb{R}^2)$ and $\mathcal{D}'(\Omega,\Lin)$ are the spaces of vector and tensor valued distributions, respectively, which are dual to the spaces of compactly supported smooth functions $\mathcal{D}(\Omega,\mathbb{R}^2)$ and $\mathcal{D}(\Omega,\Lin)$, respectively. Any locally integrable function $f$ can be associated with the distribution $T_f \in \mathcal{D}'(\Omega)$ given by $T_f (\psi)=\int_\Omega f \psi \da$ for all $\psi \in \mathcal{D}(\Omega).$ We say that a distribution $T\in \mathcal{D}'(\Omega)$ is continuous (or smooth) if there exists a continuous (or smooth)  function $f$ such that $T (\psi)=\int_\Omega f \psi \da$ for all $\psi \in \mathcal{D}(\Omega)$. 

The partial derivative of a distribution $T\in \mathcal{D}'(\Omega)$ is a distribution $\partial_i T \in \mathcal{D}'(\Omega)$ defined as $\partial_i T (\psi)=-T(\partial \psi / \partial x_i)$. The partial derivative of distributions generalizes the notion of partial derivative of differentiable functions. For a multi-index $\alpha \in \mathbb{N}^2$, where $\mathbb{N}$ is the set of non-negative integers and $\alpha=(\alpha_1,\alpha_2),$ $\partial ^\alpha T={\partial_1}^{\alpha_1}{\partial_2}^{\alpha_2} T$ with $|\alpha|=\alpha_1 + \alpha_2.$ The gradient of a distribution $T \in \mathcal{D}'(\Omega)$ is a distribution $\nabla T \in \mathcal{D}'(\Omega,\mathbb{R}^2)$ such that $\nabla T(\boldsymbol{\psi})=-T(\div \boldsymbol{\psi})$ for all $\boldsymbol{\psi}\in \mathcal{D}(\Omega,\mathbb{R}^2)$; the gradient of a distribution $\boldsymbol{T}\in \mathcal{D}'(\Omega,\mathbb{R}^2)$ is a distribution $\nabla \boldsymbol{T} \in \mathcal{D}'(\Omega,\Lin)$ given by $\nabla \boldsymbol{T}(\boldsymbol{\psi})=-\boldsymbol{T}(\div \boldsymbol{\psi})$ for all $\boldsymbol{\psi}\in \mathcal{D}(\Omega,\Lin)$. The divergence of a distribution $\boldsymbol{T} \in \mathcal{D}'(\Omega,\mathbb{R}^2)$ is a distribution $\Div \boldsymbol{T} \in \mathcal{D}'(\Omega)$ given by $\Div \boldsymbol{T}({\psi})=-\boldsymbol{T}(\nabla {\psi})$ for all ${\psi}\in \mathcal{D}(\Omega)$; the divergence of a distribution $\boldsymbol{T}\in \mathcal{D}'(\Omega,\Lin)$  is a distribution $\Div \boldsymbol{T} \in \mathcal{D}'(\Omega,\mathbb{R}^2)$ given by $\Div \boldsymbol{T}(\boldsymbol{\psi})=-\boldsymbol{T}(\nabla \boldsymbol{\psi})$ for all $\boldsymbol{\psi}\in \mathcal{D}(\Omega,\mathbb{R}^2).$ The curl of a distribution $\boldsymbol{T} \in \mathcal{D}'(\Omega,\mathbb{R}^2)$ is a distribution $\Curl \boldsymbol{T} \in \mathcal{D}'(\Omega)$ given by $\Curl \boldsymbol{T}({\psi})=-\boldsymbol{T}(\boldsymbol{e}_3\times\nabla {\psi})$ for all ${\psi}\in \mathcal{D}(\Omega)$; the curl of a distribution $\boldsymbol{T}\in \mathcal{D}'(\Omega,\Lin)$ is a distribution $\Curl \boldsymbol{T} \in \mathcal{D}'(\Omega,\mathbb{R}^2)$ such that $\langle \Curl \boldsymbol{T},\boldsymbol{a}\rangle=\Curl(\boldsymbol{T}^T \boldsymbol{a})$ for all $\boldsymbol{a}\in \mathbb{R}^2$. Note that the divergence and the curl operator associated with a distribution are distinguished by a capital initial. The Laplacian of a distribution $T\in \mathcal{D}'(\Omega)$ is a distribution $\Delta T \in \mathcal{D}'(\Omega)$ defined as $\Delta T (\psi)= T(\Delta \psi)$ for all $\psi \in \mathcal{D}(\Omega)$.

The multiplication of two arbitrary distributions is in general not well defined. The multiplication of a smooth function $f\in C^\infty (\Omega)$ with a distribution $T \in \mathcal{D}'(\Omega)$ is a distribution $fT \in \mathcal{D}'(\Omega)$ defined as $fT(\psi)=T(f\psi)$ for all $\psi \in \mathcal{D}(\Omega)$. Let $\omega$ be an open subset of $\Omega$. Given a distribution $T \in \mathcal{D}'(\Omega)$, the restriction $T|_{\omega} \in \mathcal{D}'(\omega)$ is defined as $T|_{\omega}(\psi)=T(\bar{\psi})$ for all $\psi \in \mathcal{D}(\omega)$, where $\bar{\psi}\in \mathcal{D}(\Omega)$ is such that $\bar{\psi}(\boldsymbol{x})=\psi(\boldsymbol{x})$ for all $\boldsymbol{x}\in \omega$ and $\bar{\psi}(\boldsymbol{x})=0$ for all $\boldsymbol{x} \notin \omega.$ Given a distribution $T \in \mathcal{D}'(\omega),$ $\bar{T} \in \mathcal{D}'(\Omega)$ is an extension of $T$ if $T$ is the restriction of $\bar{T}$ to $\omega$, i.e., $\bar{T}|_{\omega}=T$. A distribution is said to be continuous (or smooth) at a point $\boldsymbol{x}\in \Omega$ if there exists an open set $\omega \subset \Omega$ such that $\boldsymbol{x}\in \omega$ and the restriction $T|_{\omega}$ is continuous (or smooth). The support of a distribution $T\in \mathcal{D}'(\Omega)$, denoted by $\supp(T)$, is defined as the smallest closed set $\omega$ such that $T|_{\Omega-\omega}=0$. The singular support of a distribution $T\in\mathcal{D}'(\Omega)$, denoted by $\singsupp(T)$, is the smallest closed set $\omega$ such that $T|_{\Omega-\omega}$ is smooth. A sequence of distributions $T_n \in \mathcal{D}'(\Omega)$ converges to $T_0 \in \mathcal{D}'(\Omega)$ if $T_n (\psi) \rightarrow T_0 (\psi)$ for all $\psi \in \mathcal{D}(\Omega)$.

\subsection{Degree of distribution}
For any $0<\lambda<1,$ given $\phi \in \mathcal{D}(B_r),$ let $\tilde{\phi}_\lambda \in \mathcal{D}(B_r)$ be such that
\begin{equation}
\tilde{\phi}_\lambda(\boldsymbol{x})=\frac{1}{\lambda^2} \phi \left(\frac{\boldsymbol{x}}{\lambda}\right).
\end{equation}
The degree of a distribution $T\in \mathcal{D}'(\Omega)$ with respect to $O$ is defined as
\begin{equation}
\deg(T)=\inf\{ m\in \mathbb{R}| \lim_{\lambda \to 0} \lambda^m (T|_{B_r}) (\tilde{\phi}_\lambda)  \} - 2.
\end{equation}
The degree of a distribution $T\in \mathcal{D}'(\Omega-O)$ with respect to $O$ is similarly defined. Consider a distribution $T\in \mathcal{D}'(\Omega-O)$ such that $T(\psi)=\int_{\Omega-O} f\psi \da$ for all $\psi \in \mathcal{D}(\Omega-O)$ and  $|f(\boldsymbol{x})|\leq c_0 |\boldsymbol{x}|^m$ in a neighbourhood of $O$. Then $\deg(T)\leq -m-2$. The degree of a distribution, which is locally integrable in $\Omega-O$, describes the order of the function at point $O$. For any distribution $T\in \mathcal{D}'(\Omega),$ we have $\deg(\partial^\alpha T)<\deg(T)+|\alpha|$ for any multi index $\alpha \in \mathbb{N}^2$. For a given distribution $T\in \mathcal{D}'(\Omega-O)$ the notion of degree of distribution controls the existence and uniqueness of extension $\bar{T} \in \mathcal{D}'({\Omega})$ of $T$~\cite{brunetti2000microlocal}. For instance, for a distribution $T\in \mathcal{D}'(\Omega-O)$, with a negative degree of divergence, i.e., $\deg(T)<0$, there exists a unique extension $\bar{T} \in \mathcal{D}'(\Omega)$ such that $\bar{T}|_{\Omega-O}=T$~\cite{pandey2022point}. Several other properties of the degree of a distribution are discussed elsewhere~\cite{brunetti2000microlocal, pandey2022point}.

\subsection{Point supported distributions}
Let $\delta_O\in \mathcal{D}'(\Omega)$ represent the Dirac measure concentrated at the point $O$, i.e., $\delta_O(\psi)=\psi(O)$ for all $\psi \in \mathcal{D}(\Omega).$ We use $\mathcal{E}(\Omega)$ to represent the space of point supported distributions; a distribution $T\in \mathcal{D}'(\Omega)$ belongs to $\mathcal{E}(\Omega)$ if $\supp(T)\subset \{ O \}$. We recall the following lemma which provides the representation of any arbitrary point supported distribution as a linear combination of $\delta_O$ and its derivatives.
\begin{lemma}
\label{RepresentationELemma}
\textup{\cite[Theorem~3.2.1]{friedlander1998introduction}} For every $E\in \mathcal{E}(\Omega)$ we have the representation 
\begin{equation}
E=\sum_{\alpha \in \mathbb{N}^2, |\alpha|\leq \deg(E)} E^\alpha \partial^\alpha \delta_O, \label{representationE}
\end{equation}
with $E^\alpha \in \mathbb{R}$ given by $E^\alpha=E(v^\alpha)$, where $v^\alpha \in \mathcal{D}(\Omega)$ is such that, for any multi-index $\beta$, $\partial^\beta v^\alpha=(-1)^{|\alpha|}$ if $\alpha=\beta$ and $\partial^\beta v^\alpha=0$ if $\alpha \neq \beta$.
\end{lemma} 

\subsection{Distributional spaces}
We are interested in fields on $\Omega$ which are smooth everywhere over $\Omega$ except possibly at curve $S$ and point $O$. We model such fields as distributions whose singular support is a subset of $S \cup \{O\}$. Furthermore, we assume that the restriction of the distribution, away from any neighbourhood of $O$, can be described by bounded bulk and interfacial fields. Accordingly, we restrict ourselves to fields which can either be discontinuous across $S$ or concentrate on $S$, and can be unbounded only in the vicinity of point $O$.

We introduce two distributional subspaces $\mathcal{B}(\Omega) \subset \mathcal{D}' (\Omega)$ and $\mathcal{C}(\Omega) \subset \mathcal{D}' (\Omega)$. For any distribution $B \in \mathcal{B}(\Omega)$ there exists a piecewise smooth integrable function, with bounded derivatives in $\Omega_\epsilon -S$, $b: \Omega-\{ S\cup \{O \} \} \to \mathbb{R}$, possibly discontinuous across $S$ with $\partial S - \partial \Omega \subset \{O\}$, such that
\begin{equation}
B(\phi)=\int_{\Omega} b \phi \da \label{distB}
\end{equation}
for all $\phi \in \mathcal{D}(\Omega-O)$. The function $b$ is referred to as the bulk density of $B$. The discontinuity in $b$ is assumed to be a smooth function on $S-\{O\}$. For $\boldsymbol{x} \in S$, it is given by $\llbracket b \rrbracket (\boldsymbol{x}) = b^+ (\boldsymbol{x}) - b^- (\boldsymbol{x})$, where $b^{\pm}(\boldsymbol{x})$ are the limiting values of $b$ at $\boldsymbol{x}$ on $S$ from $\Omega^{\pm}$; the domain $\Omega^-$ is the one into which the normal $\boldsymbol{\nu}$ points. The function $\{b(\boldsymbol{x}) \}=(b^+ + b^-)/{2}$ represents the average value of $b$. Given two piecewise smooth functions $b_1$ and $b_2$, $\llbracket b_1 b_2\rrbracket=\llbracket b_1\rrbracket\{b_2\}+\llbracket b_2\rrbracket\{b_1\}$.
On the other hand, for any distribution $C \in \mathcal{C}(\Omega)$ there exists $c: S-\{O\}\to \mathbb{R}$, the line density of $C$, assumed to be a smooth bounded function on $S -\{ O \}$, such that
\begin{equation}
C(\phi)=\int_{S} c \phi \dl \label{distC}
\end{equation}
for all $\phi \in \mathcal{D}(\Omega-O)$. The spaces $\mathcal{B}(\Omega,\mathbb{R}^2)$, $\mathcal{B}(\Omega,\Lin)$, $\mathcal{C}(\Omega,\mathbb{R}^2)$, and $\mathcal{C}(\Omega,\Lin)$ can be defined in an analogous manner.

\subsection{Existence of potential fields}

According to the Poincar{\'e}'s lemma, for a simply connected open set $\Omega$, given a smooth vector field $\boldsymbol{v} \in C^\infty (\Omega,\mathbb{R}^2)$ there exists $u \in C^\infty(\Omega)$ such that $\nabla u=\boldsymbol{v}$ if and only if $\curl \boldsymbol{v}=0.$ The lemma holds true for distributional vector fields, i.e., given $\boldsymbol{V}\in \mathcal{D}'(\Omega,\mathbb{R}^2)$ there exists $U \in \mathcal{D}'(\Omega)$ satisfying $\nabla U = \boldsymbol{V}$ if and only if $\Curl \boldsymbol{V}=0$~\cite{mardare2008poincare}. In the following lemma, we recall a stronger version of the Poincar{\`e}'s lemma for distributional vector fields which belong to the distributional subspaces introduced in the preceding section. More specifically, we obtain the existence and regularity of potential functions for non-smooth piecewise discontinuous fields with concentration on a curve.

\begin{lemma}~\textup{\cite[Cor. 2.1]{pandey2020topological}}
\label{RegularityLemma}
Let $\Omega \subset \mathbb{R}^2$ be a simply connected region  and $S \subset \Omega$ be a regular oriented curve such that $\partial S-\partial \Omega\subset \{O\}$. Then,
 the condition $\Curl \boldsymbol{T}=\boldsymbol{0}$, with $\boldsymbol{T} \in \mathcal{D}'(\Omega, \mathbb{R}^2)$ and $\boldsymbol{T}(\boldsymbol{\phi})=\boldsymbol{B}(\boldsymbol{\phi})+\boldsymbol{C}(\boldsymbol{\phi})$, where $\boldsymbol{B}\in \mathcal{B}(\Omega,\mathbb{R}^2)$, $\boldsymbol{C}\in \mathcal{C}(\Omega,\mathbb{R}^2)$, and $\boldsymbol{\phi}\in \mathcal{D}(\Omega,\mathbb{R}^2)$, is equivalent to the existence of a scalar field $U \in \mathcal{B}(\Omega)$ such that $\boldsymbol{T}=\nabla U$. If $\boldsymbol{C}=\boldsymbol{0}$ then $U=\int_\Omega u \psi \da$, for all $\psi \in \mathcal{D}(\Omega-O)$, such that $u$ is a piecewise smooth scalar field continuous across the curve $S$.
\end{lemma}

We will need analogous results for symmetric tensor fields. For a simply connected open set $\Omega$, given a smooth symmetric tensor field $\boldsymbol{a} \in C^\infty (\Omega,\Sym)$, there exists $\boldsymbol{u}\in C^\infty(\Omega,\mathbb{R}^2)$ satisfying $\boldsymbol{a} = (1/2) \left(\nabla \boldsymbol{u}+(\nabla\boldsymbol{u})^T\right)$ if and only if $\curl \curl \boldsymbol{a}=\boldsymbol{0}$. More generally, given $\boldsymbol{A} \in \mathcal{D}'(\Omega,\Sym)$, there exists $\boldsymbol{U}\in \mathcal{D}'(\Omega,\mathbb{R}^2)$ satisfying $\boldsymbol{A} = (1/2) \left(\nabla \boldsymbol{U}+(\nabla\boldsymbol{U})^T\right)$ if and only if $\Curl \Curl \boldsymbol{A}=\boldsymbol{0}$~\cite{pandey2020topological}.  The following lemma establishes a stronger result for the existence and regularity of a distributional vector field given a piecewise continuous (distributional) symmetric tensor field. 

\begin{lemma}~\textup{\cite[Cor. 2.2]{pandey2020topological}}
\label{StrainCompatibilityB}
If $\Omega$ is a simply connected open subset of $\mathbb{R}^2$ and $\boldsymbol{A} \in \mathcal{B} (\Omega,\Sym)$ then
$\Curl \Curl \boldsymbol{A}=\boldsymbol{0}$
is equivalent to the existence of a vector field $\boldsymbol{U} \in \mathcal{B} (\Omega, \mathbb{R}^2)$, with $\boldsymbol{U}(\boldsymbol{\phi})=\int_\Omega \langle \boldsymbol{u},\boldsymbol{\phi}\rangle \da$ for any $\boldsymbol{\phi}\in \mathcal{D}(\Omega-O,\mathbb{R}^2)$, where $\boldsymbol{u}$ is a piecewise smooth vector field continuous across $S$, such that 
$\boldsymbol{A}= (1/2)\left({\nabla \boldsymbol{U} +(\nabla \boldsymbol{U})^T}\right)$.
\end{lemma}

Consider a distributional vector field whose singular support is a subset of $S \cup \{O\}$ and which is expressed as a sum of a piecewise smooth vector field in $\Omega-O$ and a concentration on the interfacial curve $S$. Suppose that the curl of such a vector field is point supported (at $O$);  the curl is hence expressible as a linear combination of the Dirac measure and its derivatives. In the following lemma we obtain the implications, of the curl of the vector field being point supported, on the restriction of the field to $\Omega-O$. Additionally we show that if the given field, with a point supported curl, has a negative degree of divergence then the curl of such a field is necessarily a scalar multiple of the Dirac measure. In such a case, the restriction is sufficient to obtain the curl of the distributional vector field. The lemma extends a recent result \cite[Lemma~2.9]{pandey2022point}, which was meant for vector fields smooth in $\Omega-O$, to vector fields non-smooth in $\Omega-O$.

\begin{lemma}
\label{CurlRestrictionLemma}
Let $\Omega  \subset \mathbb{R}^2$ be an open set and $\boldsymbol{V}\in \mathcal{D}'(\Omega,\mathbb{R}^2)$ such that $\boldsymbol{V}=\boldsymbol{V}_1 + \boldsymbol{V}_2$, where $\boldsymbol{V}_1\in \mathcal{B}(\Omega,\mathbb{R}^2)$, with bulk density $\boldsymbol{v}_1$, and $\boldsymbol{V}_2\in \mathcal{C}(\Omega,\mathbb{R}^2)$, with line density $\boldsymbol{v}_2$. Let $E\in \mathcal{E}(\Omega)$. 

\noindent (a) If $\Curl \boldsymbol{V}=E$ then  
\begin{subequations}
\label{CurlRestrictionImplication}
\begin{align}
 \Curl \boldsymbol{V}|_{\Omega-O}=0 ~\text{and} \label{CurlRestrictionImplication1}
 \\
  \int_{\partial B_\epsilon-S} \langle \boldsymbol{v}_1,\boldsymbol{t}\rangle \dl + \sum \langle\boldsymbol{v}_{2} (\partial B_\epsilon \cap S),\boldsymbol{\nu}\rangle = E^{(0,0)}, \label{CurlRestrictionImplication2}
 \end{align}
\end{subequations}
where the summation (denoted by $\Sigma$) is over all points of intersection of the loop $\partial B_\epsilon$ with $S$.

\noindent (b) If $\deg(\boldsymbol{V})<0$ and $E=E^{(0,0)}\delta_O$ then Equations \eqref{CurlRestrictionImplication} imply $\Curl \boldsymbol{V}=E$.
\begin{proof} 
\noindent (a) Restricting both sides of $\Curl \boldsymbol{V}=E$ to $\Omega-O$ yields $\Curl \boldsymbol{V}|_{\Omega-O}=0$. Let $\boldsymbol{V}_3=\boldsymbol{E}+(E^{(0,0)}/2\pi r)\boldsymbol{e}_\theta$, where $\boldsymbol{E}\in \mathcal{E}(\Omega,\mathbb{R}^2)$, such that $\Curl \boldsymbol{V}_3= E$~\cite[Lemma~2.8]{pandey2022point}.  Hence $\Curl (\boldsymbol{V}_3-\boldsymbol{V})=0$ which implies that there exists $U \in \mathcal{B}(B_\epsilon)$, with bulk density $u$, satisfying $\nabla U = (\boldsymbol{V}_3-\boldsymbol{V})|_{B_\epsilon }$. Consequently $\nabla u = (E^{(0,0)}/2\pi r)\boldsymbol{e}_\theta-\boldsymbol{v}_1$ in  $\Omega-\{S\cup \{O\} \}$ and $\llbracket u \rrbracket\boldsymbol{\nu}=\boldsymbol{v}_{2}$ on $\{B_\epsilon \cap S\}-\{O\}$. The expression in \eqref{CurlRestrictionImplication2} follows from integrating $\nabla u$ over the loop $\partial B_\epsilon$.

\noindent (b) If $\deg(\boldsymbol{V})<0$ then $\deg(\Curl \boldsymbol{V})<1$. Combining this with Lemma~\ref{RepresentationELemma}, and the fact that $\Curl \boldsymbol{V}$ is point supported (from \eqref{CurlRestrictionImplication1}), we obtain $\Curl \boldsymbol{V}=a \delta_O$ for some $a \in \mathbb{R}$.  If $E=E^{(0,0)}\delta_O$ then \eqref{CurlRestrictionImplication2} implies $\Curl \boldsymbol{V}=E$. 
\end{proof}
\end{lemma}

Consider a distributional symmetric tensor field $\boldsymbol{A}\in \mathcal{B}(\Omega,\Sym)$ such that $\Curl\Curl \boldsymbol{A}=0$. In the next lemma we derive the implications of such a curl free tensor field on its restriction to $\Omega-O$. We also show that the derived conditions on $\boldsymbol{A}|_{\Omega-O}$ are equivalent to $\Curl\Curl \boldsymbol{A}=0$ for $\boldsymbol{A}$ with negative degree of divergence. These results extend a recent result~\cite[Lemma~2.8]{pandey2022point}, which was meant for tensor fields smooth in $\Omega-O$, to tensor fields non-smooth in $\Omega-O$.

\begin{lemma}
\label{CurlCurlRestrictionzeroLemma}
Let $\Omega  \subset \mathbb{R}^2$ be an open set and $\boldsymbol{A}\in \mathcal{B}(\Omega,\Sym)$ a symmetric tensor field with bulk density $\boldsymbol{a}$. 

\noindent (a) $\Curl\Curl \boldsymbol{A}=0$ implies
\begin{subequations}
\label{CurlCurlRestrictionImplication}
\begin{align}
\label{CurlCurlRestrictionzero}
\Curl\Curl \boldsymbol{A}|_{\Omega-O}=0~\text{and}
\\
\label{CurlCurlLoopIntegral}
\int_{\partial B_\epsilon -S} \big( \boldsymbol{a}\boldsymbol{t} + \left( (\boldsymbol{x}-\boldsymbol{x}_0)\times\boldsymbol{e}_3\right) \langle \curl \boldsymbol{a}, \boldsymbol{t} \rangle \big) \dl + \sum \left( (\boldsymbol{x}-\boldsymbol{x}_0)\times \boldsymbol{e}_3 \right)\langle\boldsymbol\llbracket \boldsymbol{a} \rrbracket (\partial B_\epsilon \cap S),\boldsymbol{t} \otimes \boldsymbol{\nu}\rangle=0,
\end{align}
\end{subequations}
for arbitrary $\boldsymbol{x}_0 \in \mathbb{R}^2$, where the summation (denoted by $\Sigma$) is over all points of intersection of the loop $\partial B_\epsilon$ with $S$.

\noindent (b) Given $\boldsymbol{A}$ such that $\deg(\boldsymbol{A})<-2$ Equation \eqref{CurlCurlRestrictionzero} is equivalent to $\Curl\Curl \boldsymbol{A}=0$. Given $\boldsymbol{A}$ such that $\deg(\boldsymbol{A})<0$ Equations \eqref{CurlCurlRestrictionImplication} are equivalent to $\Curl\Curl \boldsymbol{A}=0$.
 \begin{proof}
 \noindent (a) $\Curl\Curl \boldsymbol{A}|_{\Omega-O}=0$ is obtained by restricting both sides of $\Curl\Curl \boldsymbol{A}=0$ to $\Omega-O$. Let $\boldsymbol{T}=\boldsymbol{A}+  \left( (\boldsymbol{x}-\boldsymbol{x}_0)\times\boldsymbol{e}_3\right) \otimes \Curl \boldsymbol{A}$. Then $\Curl \boldsymbol{T}=\left(\boldsymbol{e}_3 \times (\boldsymbol{x}-\boldsymbol{x}_0)  \right)\Curl \Curl \boldsymbol{A}$. For $\boldsymbol{A}\in \mathcal{B}(\Omega,\Sym)$ we have $\boldsymbol{T}=\boldsymbol{T}_1 + \boldsymbol{T}_2$, where $\boldsymbol{T}_1\in \mathcal{B}(\Omega,\Lin)$ and $\boldsymbol{T}_2\in \mathcal{C}(\Omega,\Lin)$. The identity $\Curl \Curl \boldsymbol{A}=0$ implies $ \Curl \boldsymbol{T}=\boldsymbol{0}$ which, using Lemma~\ref{CurlRestrictionLemma}, yields \eqref{CurlCurlLoopIntegral}.
 
\noindent (b)  Given $\Curl \Curl \boldsymbol{A}|_{\Omega-O}=0$ we have $\Curl \Curl \boldsymbol{A} \in \mathcal{E}(\Omega).$ For $\deg(\boldsymbol{A})<-2$, $\deg(\Curl \Curl \boldsymbol{A})<0$. Hence $\Curl\Curl \boldsymbol{A}=0$ is the unique extension of $\Curl \Curl \boldsymbol{A}|_{\Omega-O}=0$. On the other hand, for $\deg(\boldsymbol{A})<0$, $\deg(\Curl \Curl \boldsymbol{A})<2$. Hence, according to Lemma~\ref{RepresentationELemma}, $\Curl \Curl \boldsymbol{A}=e\delta_O +\langle \boldsymbol{b},\nabla \delta_O \rangle$ for some $e \in \mathbb{R}$ and $\boldsymbol{b}\in \mathbb{R}^2$. Then $\Curl \boldsymbol{T}=E$ with $E\in \mathcal{E}(\Omega)$ such that $E^{(0,0)}=\boldsymbol{b}\times \boldsymbol{e}_3 + e (\boldsymbol{x}_0 \times \boldsymbol{e}_3)$. According to \eqref{CurlCurlLoopIntegral} $E^{(0,0)}=0$ or, equivalently, $\boldsymbol{b}\times \boldsymbol{e}_3 + e (\boldsymbol{x}_0\times \boldsymbol{e}_3)=0$ for arbitrary $\boldsymbol{x}_0$. As a result, $\boldsymbol{b}=\boldsymbol{0}$ and $e=0$. We therefore obtain $\Curl \Curl \boldsymbol{A}=0$.
\end{proof}
\end{lemma}

In the following lemma we consider a distributional tensor field $\boldsymbol{A}$ whose singular support is a subset of $\{O\}$ and whose degree of divergence is negative. For such a field we obtain a result regarding the computation of $\Div\Div \boldsymbol{A}$.

\begin{lemma}
\label{DivDivLemma}
Let $\Omega  \subset \mathbb{R}^2$ be an open set and $\boldsymbol{A}\in \mathcal{D}'(\Omega,\Lin)$ such that $\singsupp(\boldsymbol{A})\subset\{O\}$ and $\deg(\boldsymbol{A})<0$, i.e., we can write $\boldsymbol{A}$ in terms of a smooth $\boldsymbol{a}$ as  $\boldsymbol{A}(\boldsymbol{\phi})=\int_\Omega \langle \boldsymbol{a},\boldsymbol{\phi} \rangle \da$ for all $\boldsymbol{\phi}\in \mathcal{D}(\Omega,\Lin)$. If $\Div\Div \boldsymbol{A}\in \mathcal{E}(\Omega),$ then $\Div \Div \boldsymbol{A}=e\delta_O + \langle\boldsymbol{b},\nabla\delta_O \rangle$ where $e\in \mathbb{R}$ and $\boldsymbol{b}\in \mathbb{R}^2$ such that
\begin{equation}
\int_{\partial B_\epsilon} \left(\boldsymbol{a}\boldsymbol{\nu}-(\boldsymbol{x}-\boldsymbol{x}_0)\langle\div \boldsymbol{a},\boldsymbol{\nu} \rangle \right) \dl=-e\boldsymbol{x}_0 +\boldsymbol{b}
\end{equation} 
for all $\boldsymbol{x}_0\in \mathbb{R}^2.$
\begin{proof}
Given $\deg(\boldsymbol{A})<0$ we have $\deg(\Div\Div \boldsymbol{A})<2$. Hence there exist $e\in \mathbb{R}$ and $\boldsymbol{b}\in \mathbb{R}^2$ satisfying $\Div \Div \boldsymbol{A}=e\delta_O + \langle\boldsymbol{b},\nabla\delta_O \rangle.$ Let $\boldsymbol{T}=\boldsymbol{A}-(\boldsymbol{x}-\boldsymbol{x}_0)\otimes \Div \boldsymbol{A}.$ Then $\Div \boldsymbol{T}=\boldsymbol{x}(\Div\Div \boldsymbol{A})$. Accordingly, $\Div \boldsymbol{T}=E\in \mathcal{E}(\Omega,\mathbb{R}^2)$ with $E^{0,0}=-e\boldsymbol{x}_0 +\boldsymbol{b}.$ We use \textup{\cite[Lemma 2.9(b)]{pandey2022point}} to establish the result.
\end{proof}
\end{lemma}

The next lemma establishes the existence of potential functions for distributional symmetric tensor fields which are curl and divergence free. We also discuss the regularity of these potential fields given the regularity of the symmetric tensor fields. In particular we establish the interfacial continuity of the potential field for a given regularity of the symmetric tensor field. Noticeably,  the Hessian $\nabla\nabla U\in \mathcal{D}'(\Omega,\Sym)$ of a continuous potential field $U \in \mathcal{D}'(\Omega)$ can include a concentration on an interfacial curve. This can be contrasted with Lemma~\ref{StrainCompatibilityB} where the symmetric part of the gradient of a continuous vector field necessarily does not concentrate on the interface. This is because the Hessian, unlike the symmetric part of the gradient, involves second derivatives which can concentrate for a continuous but piecewise differentiable field.
\begin{lemma}
\label{SymFieldsAiryHessian}
Let $\Omega \subset \mathbb{R}^2$ be a simply connected region  and $S \subset \Omega$ be a regular oriented curve such that $\partial S-\partial \Omega\subset \{O\}$.

\noindent (a) Given $\boldsymbol{A} \in \mathcal{D}'(\Omega, \Sym)$, $\Curl \boldsymbol{A}=\boldsymbol{0}$ if and only if there exists $U \in \mathcal{D}'(\Omega)$ such that $\nabla \nabla U =\boldsymbol{A}.$ If $\boldsymbol{A}=\boldsymbol{A}_1 + \boldsymbol{A}_2$, such that $\boldsymbol{A}_1\in \mathcal{B}(\Omega,\Sym)$ and $\boldsymbol{A}_2 \in \mathcal{C}(\Omega,\Sym)$, then $\Curl \boldsymbol{A}=\boldsymbol{0}$ if and only if there exists  $U \in \mathcal{B}(\Omega)$, where $U(\psi)=\int_\Omega u\psi \da$, for all $\psi \in \mathcal{D}(\Omega-O)$, such that  $u$ is a piecewise smooth scalar field continuous across the curve $S$, satisfying $\nabla \nabla U=\boldsymbol{A}$.

\noindent (b) Given $\boldsymbol{A} \in \mathcal{D}'(\Omega, \Sym)$, $\Div \boldsymbol{A}=\boldsymbol{0}$ if and only if there exists $U \in \mathcal{D}'(\Omega)$ such that $\mathbb{A} \nabla \nabla U =\boldsymbol{A}.$ If $\boldsymbol{A}=\boldsymbol{A}_1 + \boldsymbol{A}_2$, such that $\boldsymbol{A}_1\in \mathcal{B}(\Omega,\Sym)$ and $\boldsymbol{A}_2 \in \mathcal{C}(\Omega,\Sym)$, then $\Div \boldsymbol{A}=\boldsymbol{0}$ if and only if there exists $U \in \mathcal{B}(\Omega)$, where $U(\psi)=\int_\Omega u\psi \da$, for all $\psi \in \mathcal{D}(\Omega-O)$, such that $u$ is a piecewise smooth scalar field continuous across the curve $S$, satisfying $\mathbb{A}\nabla \nabla U = \boldsymbol{A}$.
\begin{proof}
\noindent (a) $\Curl \boldsymbol{A}=\boldsymbol{0}$ implies that there exists $\boldsymbol{V} \in \mathcal{D}'(\Omega,\mathbb{R}^2)$ such that $\boldsymbol{A}=\nabla \boldsymbol{V}$. Furthermore if $\boldsymbol{A}$ is symmetric then $\Curl \boldsymbol{V}=0$ which yields the existence of ${U} \in \mathcal{D}'(\Omega)$ such that $\nabla \nabla U = \boldsymbol{A}.$ The given regularity of $\boldsymbol{A}$ implies that $\boldsymbol{V} \in \mathcal{B}(\Omega, \mathbb{R}^2)$. The required regularity of $U$ then follows from $\nabla U =\boldsymbol{V}$.

\noindent (b) The proof follows from that in (a) once we identify $\Div \boldsymbol{A}=\Curl (\boldsymbol{e}_3 \times \boldsymbol{A})$.
\end{proof}
\end{lemma}

\subsection{The Monge-Amp{\`e}re bracket} \label{mabrack}

For any open set $\omega \subset \Omega,$ given two smooth scalar fields $b_1 \in C^\infty (\omega)$ and $b_2 \in C^\infty(\omega),$ the Monge-Amp{\`e}re bracket, $[b_1,b_2]\in  C^\infty (\omega)$ is defined as $[b_1,b_2]=\langle\mathbb{A}\nabla \nabla b_1, \nabla\nabla b_2\rangle.$ The Monge-Amp{\`e}re bracket is a symmetric operator, i.e.,  $[b_1,b_2]=[b_2,b_1]$, and satisfies the identity $\curl \curl (\nabla b_1 \otimes \nabla b_2)=-[b_1,b_2]$. We generalize the Monge-Amp{\`e}re bracket to distributions (i.e., to non-smooth scalar fields) using one of the following:

\begin{enumerate}[align = left]
\item[\mylabel{a1}{(A1)}] Consider two scalar fields such that one is smooth and the other one is a distribution, i.e., $B_1 \in C^\infty (\Omega)$ and $B_2 \in \mathcal{D}'(\Omega)$. In such a case we can use the notion of multiplication of a smooth function and a distribution to argue that $\nabla B_1 \otimes \nabla B_2 \in \mathcal{D}'(\Omega)$ and $[B_1,B_2] =\langle\mathbb{A}\nabla \nabla B_1, \nabla\nabla B_2\rangle \in \mathcal{D}'(\Omega)$.

\item[\mylabel{a2}{(A2)}] Consider two distributional scalar fields $B_1 \in \mathcal{B}(\omega)$ and $B_2 \in \mathcal{B}(\omega)$, with bulk densities $b_1$ and $b_2$, respectively, satisfying three assumptions: (i) $b_1$ and $b_2$ are piecewise smooth but continuous across $S$, (ii) $\deg(\nabla B_1)<0$ and $\deg(\nabla B_2)<0$, and (iii) $\deg(\nabla B_1|_{\omega-O} \otimes \nabla B_2|_{\omega-O})<0$, where $\nabla B_1|_{\omega-O} \otimes \nabla B_2|_{\omega-O} \in \mathcal{D}'(\omega-O,\Lin)$ such that $\nabla B_1|_{\omega-O} \otimes \nabla B_2|_{\omega-O}(\boldsymbol{\psi})=\int_{\Omega-O} \langle \nabla b_1\otimes \nabla b_2, \boldsymbol{\psi} \rangle\da$ for all $\boldsymbol{\psi}\in \mathcal{D}(\Omega-O,\Lin)$. The distributions $B_1$ and $B_2$ have, in general, intersecting singular support which prohibits us to directly use the multiplication of a distribution and a smooth function. Assumption (i) implies that $\nabla B_1 \in \mathcal{B}(\omega,\mathbb{R}^2)$ and $\nabla B_2 \in \mathcal{B}(\omega,\mathbb{R}^2)$; accordingly, $\nabla B_1|_{\omega-O}$ and $\nabla B_2|_{\omega-O}$ are piecewise continuous vector valued functions defined uniquely by $\nabla b_1$ and $\nabla b_2$.  Assumption (ii) implies that $\nabla B_1$ and $\nabla B_2$ are uniquely defined given $\nabla B_1|_{\omega-O}$ and $\nabla B_2|_{\omega-O}$. Consequently, (i) and (ii) simultaneously imply that $\nabla B_1$ and $\nabla B_2$ are uniquely defined through piecewise continuous functions $\nabla b_1$ and $\nabla b_2$, respectively. The dyadic product $\nabla B_1|_{\omega-O} \otimes \nabla B_2|_{\omega-O}$ then gets defined as the point wise dyadic product of $\nabla b_1$ and $\nabla b_2$. Finally, due to assumption (iii), we can introduce $\nabla B_1 \otimes \nabla B_2\in \mathcal{D}'(\omega,\Lin)$ as the unique extension of $\nabla B_1|_{\omega-O} \otimes \nabla B_2|_{\omega-O}$.  The generalized Monge-Amp{\`e}re bracket $[B_1,B_2]$ is then defined as $-\Curl \Curl (\nabla B_1 \otimes \nabla B_2)$. 

\item[\mylabel{a2p}{(A2')}] Consider a distributional scalar field $B\in \mathcal{B}(\Omega)$, with bulk density $b$, satisfying three assumptions: (i) $b$ is piecewise smooth but continuous across $S$, (ii) $\deg(\nabla B) <0$, and (iii) $\deg(\nabla B|_{\Omega-O} \otimes \nabla B|_{\Omega-O})<0$. The implications of these assumptions are as described in \ref{a2} above. The field $\nabla B \otimes \nabla B \in \mathcal{D}'(\Omega,\Lin)$ is introduced as the unique extension of $\nabla B|_{\Omega-O} \otimes \nabla B|_{\Omega-O}$ and the generalized Monge-Amp{\`e}re bracket $[B,B]$ is defined as $-\Curl \Curl (\nabla B \otimes \nabla B)$. 
\end{enumerate}
Essentially, we are able to generalize the notion of Monge-Amp{\`e}re bracket for certain distributional fields whenever we are able to clearly define the tensor product of their gradients. If both $B_1$ and $B_2$ are smooth fields then they satisfy the assumptions in \ref{a1} and \ref{a2} and the generalized Monge-Amp{\`e}re bracket reduces to the usual Monge-Amp{\`e}re bracket for smooth fields. If both of them are twice differentiable then they satisfy the assumptions in \ref{a2} and again the generalized Monge-Amp{\`e}re bracket reduces to the usual Monge-Amp{\`e}re bracket.

The inner product of two distributional tensor fields and the determinant of a distributional tensor field are both in general not well defined. We use the generalized Monge-Amp{\`e}re bracket to discuss an exceptional situation when both of these can be defined unambiguously. We first recall a classical result.
Let $\boldsymbol{a}_1 \in C^\infty (\Omega,\Sym)$ and $\boldsymbol{a}_2 \in C^\infty (\Omega,\Sym)$ be two smooth symmetric tensor fields such that there exist fields $b_1 \in C^\infty(\Omega)$ and $b_2 \in C^\infty(\Omega)$ satisfying $\mathbb{A}\nabla \nabla b_1=\boldsymbol{a}_1$ and $\nabla \nabla b_2=\boldsymbol{a}_2$. We then have $\langle \boldsymbol{a}_1,\boldsymbol{a}_2 \rangle=[b_1,b_2]$ and $\det(\boldsymbol{a}_2)=\frac{1}{2} [b_2,b_2]$. In order to extend these identifications to distributions we consider two distributional symmetric tensor fields $\boldsymbol{A}_1 \in \mathcal{D}'(\omega,\Sym)$ and $\boldsymbol{A}_2 \in \mathcal{D}'(\omega,\Sym)$ such that there exist scalar fields $B_1 \in \mathcal{D}'(\Omega)$ and $B_2 \in \mathcal{D}'(\Omega)$ satisfying $\mathbb{A}\nabla \nabla B_1=\boldsymbol{A}_1$ and $\nabla \nabla B_2=\boldsymbol{A}_2$. Additionally, we require $B_1$ and $B_2$ to be fields which satisfy the assumptions mentioned in either \ref{a1} or \ref{a2}. For such symmetric tensor fields we define the inner product $\langle \boldsymbol{A}_1,\boldsymbol{A}_2 \rangle \in \mathcal{D}'(\omega)$ as 
\begin{equation}
\langle \boldsymbol{A}_1,\boldsymbol{A}_2 \rangle=[B_1,B_2], \label{innerprod}
\end{equation}
 where the Monge-Amp{\`e}re bracket is as introduced in \ref{a1} or \ref{a2}. On the other hand, we consider a  distributional symmetric tensor field $\boldsymbol{A}\in \mathcal{D}' (\Omega,\Sym)$ such that there exist a scalar field $B \in \mathcal{D}'(\Omega)$ satisfying  $\nabla \nabla B=\boldsymbol{A}$. Additionally, we require $B$ to be a field which satisfies the assumptions mentioned in \ref{a2p}. The determinant of $\boldsymbol{A}$ can then be generalized as 
 \begin{equation}
 \Det(\boldsymbol{A})=\frac{1}{2} [B,B], \label{distdet}
\end{equation}
where the Monge-Amp{\`e}re bracket is as introduced in \ref{a2p}; note that we are representing the distributional determinant as $\Det$ (in comparison to the usual $\det$). Equation \eqref{distdet} provides a definition of the distributional determinant of symmetric tensor fields which are given in terms of the distributional Hessian of certain scalar fields. In this way, we are able to calculate the determinant of such symmetric tensor fields even if they are non integrable and  develop concentration on a smooth curve.

\begin{remark} (Distributional Jacobian)
Given a twice differentiable vector field $\boldsymbol{u}=u_1\boldsymbol{e}_1+u_2\boldsymbol{e}_2$, the determinant of its gradient (the Jacobian) can be expressed in terms of a divergence as $\det(\nabla \boldsymbol{u})=\div(u_1 (\nabla u_2 \times \boldsymbol{e}_3)).$ For distributional tensor fields, which are expressible as the gradient of a vector field, a distributional determinant can be defined using the distributional divergence in this identity. The distributional Jacobian enjoys certain weak continuity properties when expressed in this distributional divergence form~\cite{muller1990det}; it develops singular regions under weaker regularity conditions in excess to the point wise Jacobian~\cite{muller1993singular}.  Analogously, the distributional determinant in \eqref{distdet} develops singular regions in excess to the point wise determinant. Elsewhere in this paper, these excess quantities are calculated in a strong form for the case when the Hessian field concentrates on a curve.
\end{remark}

\begin{remark} (Monge-Amp{\`e}re bracket as a weak sequential limit) In this remark, we show that the generalization of the Monge-Amp{\`e}re bracket for certain distributional fields is the weak sequential limit of the Monge-Amp{\`e}re bracket of smooth fields. We consider two fields $B_1$ and $B_2$, both in $\mathcal{B}(\Omega)$, such that $\nabla \nabla B_2$ concentrates on the interface whereas $\nabla \nabla B_1$ is a piecewise continuous function which does not concentrate on the interface; these regularity assumptions are stronger than those considered in \ref{a2} where the Hessian of both the fields was allowed to concentrate on the interfacial curve.
In more formal terms, let $B_1\in \mathcal{B}(\Omega)$ with bulk density $b_1$ be such that $\nabla\nabla B_1 \in \mathcal{B}(\Omega,\Sym)$ is square integrable, i.e., $\nabla\nabla b_1 \in L^2(\Omega,\Sym)$, and $B_2 \in \mathcal{B}(\Omega)$ with bulk density $b_2$ be such that $\nabla B_2 \in \mathcal{B}(\Omega,\Sym)$ is square integrable, i.e., $\nabla b_1 \in L^2(\Omega,\mathbb{R}^2)$.
For sequences ${b_1}^n \in C^\infty (\Omega)$ and ${b_2}^n \in C^\infty(\Omega)$, such that $\nabla \nabla {b_1}^n \rightharpoonup \nabla \nabla {b_1}$ weakly in $L^2(\Omega,\Sym)$ and $\nabla {b_2}^n \rightharpoonup  \nabla {b_2}$ weakly in $L^2(\Omega,\mathbb{R}^2)$, $[{b_1}^n,{b_2}^n] \to [B_1,B_2]$ in the sense of distributions.
Indeed, $\curl(\nabla {b_1}^n \otimes \nabla {b_2}^n)=(\langle -\partial_{12} {b_1}^n \boldsymbol{e}_1 +\partial_{11} {b_1}^n \boldsymbol{e}_2, \nabla {b_2}^n  \rangle) \boldsymbol{e}_1+(\langle -\partial_{22} {b_1}^n \boldsymbol{e}_1 +\partial_{12} {b_1}^n \boldsymbol{e}_2, \nabla {b_2}^n  \rangle) \boldsymbol{e}_2.$ We note that $\curl(\nabla {b_2}^n)=0$ and $\div(-\partial_{12} {b_1}^n \boldsymbol{e}_1 +\partial_{11} {b_1}^n \boldsymbol{e}_2)=\div (-\partial_{22} {b_1}^n \boldsymbol{e}_1 +\partial_{12} {b_1}^n \boldsymbol{e}_2)=0$. The Div-Curl lemma~\cite[Theorem~5.2.1]{evans1990weak} then implies that $\curl (\nabla {b_1}^n \otimes \nabla {b_2}^n) \to \Curl (\nabla {b_1} \otimes \nabla {b_2})$ in the sense of distributions. Using $[{b_1}^n,{b_2}^n]=- \curl \curl (\nabla {b_1}^n \otimes \nabla {b_2}^n),$ we have $[{b_1}^n,{b_2}^n] \to -\Curl \Curl (\nabla {b_1} \otimes \nabla {b_2})$ in the sense of distributions. \label{maseqlim}
\end{remark}

\begin{remark}(The inner product of distributional tensor fields) We were able to define an inner product of distributional tensor fields $\boldsymbol{A}_1 \in \mathcal{D}'(\Omega,\Sym)$ and $\boldsymbol{A}_2 \in \mathcal{D}'(\Omega,\Sym)$ in \eqref{innerprod} when the tensor fields were given in terms of potential fields $B_1\in \mathcal{D}'(\Omega)$ and  $B_2 \in \mathcal{D}'(\Omega)$, satisfying the assumptions in either \ref{a1} or \ref{a2}, such that $\boldsymbol{A}_1=\mathbb{A}\nabla \nabla B_1$ and $\boldsymbol{A}_2=\nabla\nabla B_2$. We now show that $\langle \boldsymbol{A}_1,\boldsymbol{A}_2 \rangle$ is in fact independent of the choice of $B_1$ and $B_2$. In other words, given another pair of potential fields $B_1' \in \mathcal{D}'(\Omega)$ and $B_2' \in \mathcal{D}'(\Omega)$, satisfying the assumptions in either \ref{a1} or \ref{a2}, such that $\mathbb{A}\nabla \nabla B_1' = \boldsymbol{A}_1$ and $\nabla \nabla B_2' = \boldsymbol{A}_2$, we have $\Curl \Curl (\nabla B_1\otimes \nabla B_2)=\Curl \Curl (\nabla B_1' \otimes \nabla B_2')$.
Indeed, $\mathbb{A}\nabla \nabla B_1=\mathbb{A}\nabla \nabla B_1'$ and $\nabla \nabla B_2=\nabla \nabla B_2'$ implies that there exist $\boldsymbol{c}_1 \in \mathbb{R}^2$ and $\boldsymbol{c}_2 \in \mathbb{R}^2$ such that $\nabla B_1=\nabla B_1' + \boldsymbol{c}_1$ and $\nabla B_2=\nabla B_2' + \boldsymbol{c}_2$. Noting $\Curl \Curl (\boldsymbol{c}_1 \otimes \nabla B_2) = 0$ and $\Curl \Curl (\nabla B_1\otimes \boldsymbol{c}_2)=0$ implies the required result.
\end{remark}

\subsection{Identities} 

In this section, we collect a set of identities which will be useful for explicit computation of local forms of certain fields away from the singular point $O$. The following pair of identities computes the local densities corresponding to $\Curl \Curl \boldsymbol{A}$ for $\boldsymbol{A}\in \mathcal{B}(\Omega,\Lin)$. These computations are immediately applicable in calculating local densities for the generalized Monge-Amp{\`e}re bracket, the inner product, and the determinant. 

 \noindent \begin{id}
\label{CurlCurlFormulas}
Let $\boldsymbol{A}\in \mathcal{B}(\Omega,\Lin)$ such that $\boldsymbol{A}(\boldsymbol{\phi})=\int_\Omega  \langle \boldsymbol{a}, \boldsymbol{\phi} \rangle \da$ for all $\boldsymbol{\phi} \in \mathcal{D}(\Omega,\Lin)$.

\noindent (a)  For $\psi \in \mathcal{D}(\Omega-O)$,
\begin{equation}
\label{CurlCurl}
\Curl \Curl \boldsymbol{A} (\psi) = \int_{\Omega-S} \left( \curl \curl \boldsymbol{a} \right) \psi \da + \int_S  \langle \llbracket \boldsymbol{a}\rrbracket, \boldsymbol{t}\otimes \boldsymbol{t}  \rangle \frac{\partial \psi}{\partial \nu}\dl+ 
 \int_S  \big( \langle\llbracket \nabla \boldsymbol{a}\rrbracket, \boldsymbol{r}  \rangle  + k \langle \llbracket \boldsymbol{a}\rrbracket, \boldsymbol{s} \rangle \big) \psi \dl, 
\end{equation}
where ${\partial \psi}/{\partial \nu} = \langle \nabla \psi, \boldsymbol{\nu} \rangle$, 
\begin{equation}
\boldsymbol{r} = (\boldsymbol{t}\otimes \boldsymbol{\nu}\otimes \boldsymbol{t}  -  \boldsymbol{t}\otimes \boldsymbol{t}\otimes \boldsymbol{\nu} +\boldsymbol{\nu}\otimes \boldsymbol{t}\otimes \boldsymbol{t}), ~\text{and}~ \boldsymbol{s} = (\boldsymbol{\nu}\otimes \boldsymbol{\nu} - \boldsymbol{t}\otimes \boldsymbol{t}). \label{defrs}
\end{equation}

\noindent (b) If $\boldsymbol{a}=\nabla a_1 \otimes \nabla a_2$ in $\Omega -S$, where $a_1, a_2$ are piecewise smooth scalar maps continuous across the interface $S$, then, for $\psi \in \mathcal{D}(\Omega-O)$,
\begin{equation}
\label{MongeAmpereInterface}
\Curl \Curl \boldsymbol{A} (\psi) =- \int_{\Omega-S} [a_1,a_2] \psi \da+ \int_S \big(\langle \{\nabla\nabla a_1\} , \boldsymbol{t}\otimes \boldsymbol{t} \rangle  \langle \llbracket \nabla a_2\rrbracket,\boldsymbol{\nu} \rangle   + \langle \{\nabla\nabla a_2\} , \boldsymbol{t}\otimes \boldsymbol{t} \rangle  \langle \llbracket \nabla a_1\rrbracket,\boldsymbol{\nu} \rangle  \big) \psi \dl.
\end{equation}
\begin{proof}
\noindent (a) The proof follows immediately from \cite[Id. 2.3]{pandey2020topological}.

\noindent (b) Identity \eqref{MongeAmpereInterface} follows from \eqref{CurlCurl} after noting the following: $\llbracket\langle\nabla a_1, \boldsymbol{t}\rangle\rrbracket=0,$ $\llbracket\langle\nabla a_2, \boldsymbol{t}\rangle\rrbracket=0,$ $\llbracket \langle \nabla\nabla a_1,\boldsymbol{t}\otimes \boldsymbol{t}\rangle \rrbracket=-k\llbracket \langle\nabla a_1,\boldsymbol{\nu} \rangle \rrbracket$, $\llbracket \langle \nabla\nabla a_2,\boldsymbol{t}\otimes \boldsymbol{t}\rangle \rrbracket=-k\llbracket \langle\nabla a_2,\boldsymbol{\nu} \rangle \rrbracket$, 
$\llbracket \langle \nabla(\nabla a_1 \otimes \nabla a_2), \boldsymbol{t}\otimes \boldsymbol{\nu} \otimes \boldsymbol{t} \rangle\rrbracket=\llbracket \langle\nabla\nabla a_1,\boldsymbol{t}\otimes \boldsymbol{t} \rangle \langle\nabla a_2,\boldsymbol{\nu} \rangle\rrbracket+ \llbracket\langle\nabla\nabla a_2,\boldsymbol{\nu}\otimes \boldsymbol{t} \rangle \langle\nabla a_1,\boldsymbol{t} \rangle\rrbracket$,
 $\llbracket \langle \nabla(\nabla a_1 \otimes \nabla a_2), \boldsymbol{\nu}\otimes \boldsymbol{t} \otimes \boldsymbol{t} \rangle \rrbracket=\llbracket\langle\nabla\nabla a_1,\boldsymbol{\nu}\otimes \boldsymbol{t} \rangle \langle\nabla a_2,\boldsymbol{t} \rangle \rrbracket+ \llbracket\langle\nabla\nabla a_2,\boldsymbol{t}\otimes \boldsymbol{t} \rangle \langle\nabla a_1,\boldsymbol{\nu} \rangle\rrbracket$, and
$\llbracket \langle \nabla(\nabla a_1 \otimes \nabla a_2), \boldsymbol{t}\otimes \boldsymbol{t} \otimes \boldsymbol{\nu} \rangle \rrbracket = \llbracket \langle\nabla\nabla a_1,\boldsymbol{t}\otimes \boldsymbol{\nu} \rangle \langle\nabla a_2,\boldsymbol{t} \rangle \rrbracket + \llbracket \langle\nabla\nabla a_2,\boldsymbol{t}\otimes \boldsymbol{\nu} \rangle \langle\nabla a_1,\boldsymbol{t} \rangle\rrbracket$.
\end{proof}
\end{id}

In the following identity the local densities corresponding to a distributional Laplacian are obtained.

 \noindent \begin{id}
\label{LaplacianFormula}
Let ${A}\in \mathcal{D}'(\Omega)$ such that $A=A_1+A_2$, where $A_1 \in \mathcal{B}(\Omega)$ with bulk density $a_1$ and $A_2 \in \mathcal{C}(\Omega)$ with line density $a_2$, i.e., ${A}({\psi})=\int_\Omega   {a_1}{\psi} \da+\int_S  a_2 \psi \dl$ for all ${\psi} \in \mathcal{D}(\Omega-O)$. Then, for $\psi \in \mathcal{D}(\Omega-O)$,

\begin{equation}
\label{Laplacian}
\Delta A (\psi)=\int_{\Omega-S} \Delta a_1 \psi \da - \int_S \left(\langle\llbracket \nabla a_1\rrbracket,\boldsymbol{\nu}\rangle - \frac{d^2 a_2}{ds^2} \right)\psi \dl + \int_S \left( \llbracket a_1 \rrbracket -ka_2 \right) \frac{\partial \psi}{\partial \nu} \dl +  \int_S a_2 \langle\nabla\nabla \psi,\boldsymbol{\nu}\otimes \boldsymbol{\nu} \rangle \dl
\end{equation}
\begin{proof}
Identity \eqref{Laplacian} can be proved as a straightforward application of \cite[Id. 2.1]{pandey2020topological} and\cite[Id. 2.2]{pandey2020topological}.
\end{proof}
\end{id}
 
\section{Inhomogeneous von K{\'a}rm{\'a}n equations with singular fields}

We begin by recalling the inhomogeneous von K{\'a}rm{\'a}n plate equations with smooth fields, with sources of inhomogeneity given in terms of incompatibility (arising out of defect densities) and metric anomaly fields, and then, in rest of the section, generalize the von K{\'a}rm{\'a}n equations to fields which are singular at a point and a curve in the plate domain. This generalization is the central contribution of the paper.

\subsection{Inhomogeneous von K{\'a}rm{\'a}n equations with smooth fields} \label{inhomovksmooth} The following derivation of smooth von K{\'a}rm{\'a}n equations follows our recent work~\cite{manish1, manish2}. Let $\Omega \subset \mathbb{R}^2$ be a simply connected open set representing the plate domain. Given an in-plane displacement field $\boldsymbol{u} \in C^\infty (\Omega,\mathbb{R}^2)$ and a transverse displacement field $w \in C^\infty (\Omega)$, with respect to a planar reference configuration, the smooth stretching and bending strains within von K{\'a}rm{\'a}n plate kinematics are of the form 
\begin{equation}
\boldsymbol{e} = \frac{1}{2}\left( \nabla \boldsymbol{u} +\nabla \boldsymbol{u}^T \right) + \frac{1}{2} \nabla w \otimes \nabla w ~\text{and}~ \boldsymbol{\lambda} = \nabla \nabla w, \label{vksmooth1}
\end{equation}
respectively. On the other hand, given strain fields  $\boldsymbol{e}\in C^\infty (\Omega,\Sym)$ and $\boldsymbol{\lambda}\in C^\infty (\Omega,\Sym)$, the compatibility conditions which are necessary and sufficient for the existence of the displacement fields, such that \eqref{vksmooth1} are satisfied, are 
\begin{equation}
\curl \boldsymbol{\lambda}=\boldsymbol{0}~\text{and}~\curl \curl \boldsymbol{e} + \det (\boldsymbol{\lambda})=0. \label{vksmooth2}
\end{equation}
The strain fields are additively decomposed into elastic ($\boldsymbol{e}^e, \boldsymbol{\lambda}^e$) and plastic ($\boldsymbol{e}^p, \boldsymbol{\lambda}^p$) parts, all smooth, as $\boldsymbol{e} = \boldsymbol{e}^e + \boldsymbol{e}^p$ and $\boldsymbol{\lambda} = \boldsymbol{\lambda}^e + \boldsymbol{\lambda}^p$. It is worth noting that this decomposition is more general than what is used in linear elasticity theories since the decomposition of stretching strain and of bending strain are at different order of magnitudes. The elastic and plastic strain fields are incompatible in the sense that they do not necessarily satisfy the compatibility equations given above. In other words, there exist incompatibility fields $\eta_1 \in C^\infty (\Omega)$ and $\boldsymbol{\eta}_2 \in C^\infty (\Omega,\mathbb{R}^2)$ such that 
\begin{equation}
\curl \boldsymbol{\lambda}^p=\boldsymbol{\eta}_2~\text{and}~\curl \curl \boldsymbol{e}^p + \det (\boldsymbol{\lambda}^p)=\eta_1. \label{vksmooth3}
\end{equation}
The incompatibility fields can be interpreted in terms of defect densities. The equilibrium equations for a von K{\'a}rm{\'a}n plate are given in terms of an in-plane stress tensor $\boldsymbol{\sigma}\in C^\infty (\Omega,\Sym)$ and a moment tensor $\boldsymbol{m}\in C^\infty (\Omega,\Sym)$ as 
\begin{equation}
\div \boldsymbol{\sigma}= \boldsymbol{0}~\text{and}~\div \div \boldsymbol{m}- \langle \boldsymbol{\sigma},\boldsymbol{\lambda}\rangle=f,\label{vksmooth4}
\end{equation}
where $f \in C^\infty (\Omega)$ is the transverse body force field. The equilibrium equation \eqref{vksmooth4}$_1$ is equivalent to the existence of a stress function $\phi \in C^\infty (\Omega)$ such that $\boldsymbol{\sigma}=\mathbb{A}\nabla \nabla \phi$. The remaining ingredients of the theory are the constitutive relations for a materially uniform isotropic elastic plate: 
\begin{equation}
\boldsymbol{\sigma}=\frac{E}{1-\nu^2}\big( (1-\nu)\boldsymbol{e}^e+ \nu \tr(\boldsymbol{e}^e) \boldsymbol{I} \big)~\text{and}~
 \boldsymbol{m}=D\big( (1-\nu)\boldsymbol{\lambda}^e+ \nu \tr(\boldsymbol{\lambda}^e) \boldsymbol{I} \big), \label{vksmooth5}
\end{equation}
 where $E$, $D$, and $\nu$ are the stretching modulus, bending modulus, and Poisson ratio of the 2D elastic surface. The first von K{\'a}rm{\'a}n equation is obtained by first substituting $\boldsymbol{e}$ in terms of $\boldsymbol{e}^e$ and $\boldsymbol{e}^p$ into \eqref{vksmooth2}$_2$, then replacing $\boldsymbol{e}^e$ in terms of the stress function, and finally using \eqref{vksmooth3}$_2$ to replace $\boldsymbol{e}^p$ to obtain 
\begin{equation}
\frac{1}{E}\Delta^2 \phi +\frac{1}{2} [w,w]=-\eta_1 + \det({\boldsymbol\lambda}^p)  ~ \text{in}~\Omega. \label{vksmooth6}
\end{equation}
The second von K{\'a}rm{\'a}n equation is obtained by first replacing $\boldsymbol{m}$ in terms of ${\boldsymbol\lambda}^e$, using \eqref{vksmooth5}$_2$, and $\boldsymbol{\sigma}$ in terms of the stress function in \eqref{vksmooth4}$_2$, and then substituting ${\boldsymbol\lambda}^e$ in terms of ${\boldsymbol\lambda}$ (hence $w$) and ${\boldsymbol\lambda}^p$ as
\begin{equation}
D\Delta^2 w - [\phi,w]=f + D \big( (1-\nu)\div \div {\boldsymbol{\lambda}^p} +\nu \Delta \tr({\boldsymbol{\lambda}^p})\big)  ~ \text{in}~\Omega. \label{vksmooth7}
\end{equation}
The two incompatible von K{\'a}rm{\'a}n equations with smooth fields, given in \eqref{vksmooth6} and \eqref{vksmooth7}, when combined with suitable boundary conditions, can be used for the evaluation of stress field and the transverse displacement field of the plate for a given smooth distribution of defects and metrical anomalies.  

\subsection{Generalized compatibility conditions} \label{weakcomp}
In this section, we generalize the strain displacement relations \eqref{vksmooth1} and the compatibility relations \eqref{vksmooth2} to a distributional form (and the equivalent local forms) for non-smooth fields over $\Omega$. We consider the in-plane displacement $\boldsymbol{u}:\Omega \to \mathbb{R}^2$ and the transverse displacement  ${w}:\Omega \to \mathbb{R}$ such that they are piecewise smooth but continuous across the interface $S$ and bounded in the domain $\Omega$. Their derivatives are allowed to be unbounded at $O$. We define distributional displacement fields $\boldsymbol{U}\in \mathcal{B}(\Omega,\mathbb{R}^2)$, given by $\boldsymbol{U}(\boldsymbol{\psi})=\int_\Omega \left\langle \boldsymbol{u},\boldsymbol{\psi}\right\rangle \da$ for all $\boldsymbol{\psi} \in \mathcal{D}(\Omega,\mathbb{R}^2)$, and ${W}\in \mathcal{B}(\Omega)$, given by ${W}({\psi})=\int_\Omega  {w}{\psi} \da$ for all ${\psi} \in \mathcal{D}(\Omega)$. We restrict ourselves to ${w}$ such that $W$ satisfies the assumptions in \ref{a2p} (i.e., $\deg(\nabla W)<0$ and $\deg(\nabla W|_{\Omega-O} \otimes \nabla W|_{\Omega-O})<0$); consequently, we can define $\nabla W \otimes \nabla W \in \mathcal{D}'(\Omega,\Sym)$ as the unique extension of $\nabla W|_{\Omega-O} \otimes \nabla W|_{\Omega-O}$ such that $\deg(\nabla W \otimes \nabla W)=\deg(\nabla W|_{\Omega-O} \otimes \nabla W|_{\Omega-O})$. The distributional stretching strain $\boldsymbol{E}\in\mathcal{D}'(\Omega,\Sym)$ and the distributional bending strain $\boldsymbol{\Lambda}\in \mathcal{D}'(\Omega,\Sym)$ are introduced as
\begin{equation}
\label{StrainsDistributionalDefinition}
\boldsymbol{E}=\frac{1}{2}\left( \nabla \boldsymbol{U} + \nabla \boldsymbol{U} ^T\right) + \frac{1}{2} \nabla W \otimes \nabla W~\text{and}~\boldsymbol{\Lambda}=\nabla \nabla W,
\end{equation}
respectively. 
Recalling  \cite[Id.~2.1]{pandey2020topological}, it immediately follows that there exists a bulk stretching strain $\boldsymbol{e}:\Omega-\{S\cup \{O\}\}\to \Sym$, satisfying $\boldsymbol{E}(\boldsymbol{\psi})=\int_\Omega\left\langle \boldsymbol{e},\boldsymbol{\psi} \right\rangle \da$ for all $\boldsymbol{\psi} \in \mathcal{D}(\Omega-O,\Lin)$, and a bulk bending strain $\boldsymbol{\lambda}:\Omega-\{S\cup\{O\}\}\to \Sym$ and a concentrated (on $S$) bending strain $\boldsymbol{\gamma} : S-\{O\}\to \Sym$, satisfying $\boldsymbol{\Lambda}(\boldsymbol{\psi})=\int_\Omega\left\langle \boldsymbol{\lambda},\boldsymbol{\psi} \right\rangle \da + \int_S\left\langle \boldsymbol{\gamma},\boldsymbol{\psi} \right\rangle \dl$ for all $\boldsymbol{\psi} \in \mathcal{D}(\Omega-O,\Lin)$, 
such that
\begin{subequations}
\label{DefinitionofStrains}
\begin{align}
\label{DefinitionStretchingStrainBulk}
\boldsymbol{e} = \frac{1}{2}\left( \nabla \boldsymbol{u} +\nabla \boldsymbol{u}^T \right) + \frac{1}{2} \nabla w \otimes \nabla w  ~\text{in}~ \Omega-\{S\cup\{O\}\},
\\
\label{DefinitionCurvatureStrainBulk}
 \boldsymbol{\lambda} = \nabla \nabla w  ~\text{in}~ \Omega-\{S\cup\{O\}\},~\text{and}
\\
\label{DefinitionCurvatureStrainConcentration}
 \boldsymbol{\gamma} = - \llbracket \nabla w \rrbracket \otimes \boldsymbol{\nu}  ~\text{on}~ S-\{O\}.
\end{align}
\end{subequations} 
Note that Equations \eqref{DefinitionofStrains} imply \eqref{StrainsDistributionalDefinition} only when $\deg(\boldsymbol{E})<0$, $\deg(\boldsymbol{\Lambda})<0$, and $\deg(W)<-2$. Also, due to the continuity of the displacement fields $\boldsymbol{u}$ and $w$ across $S$, the stretching strain does not develop any concentration on $S$ as it is defined through the first derivatives of the displacement fields; on the other hand, the bending strain does develop a concentration as it is defined in terms of the distributional Hessian of the transverse displacement. 

We now seek conditions on distributional strain fields $\boldsymbol{E}\in \mathcal{D}'(\Omega,\Sym)$ and $\boldsymbol{\Lambda}\in \mathcal{D}'(\Omega,\Sym)$ such that there exist distributional displacements $W\in \mathcal{D}'(\Omega)$ and $\boldsymbol{U} \in \mathcal{D}'(\Omega,\mathbb{R}^2)$ satisfying \eqref{StrainsDistributionalDefinition}. According to Lemma~\ref{SymFieldsAiryHessian} there exists $W\in \mathcal{D}'(\Omega)$ satisfying $\boldsymbol{\Lambda}=\nabla\nabla W$ if and only if 
\begin{equation}
\label{CurlLambdaDistributional}
\Curl \boldsymbol{\Lambda}=\boldsymbol{0}
\end{equation}
is satisfied. We require that $W$ satisfy the assumptions in \ref{a2p}. This is necessary for us to introduce a generalized notion of the determinant of $\boldsymbol{\Lambda}$. The necessary and sufficient condition, for the existence of $\boldsymbol{U} \in \mathcal{D}'(\Omega,\mathbb{R}^2)$ satisfying \eqref{StrainsDistributionalDefinition}$_1$, then follows from \cite[Cor. 2.1]{pandey2020topological} as
\begin{equation}
\label{CompatibiilityDistributional}
\Curl \Curl \boldsymbol{E} + \Det (\boldsymbol{\Lambda})=0,
\end{equation} 
where the determinant is to be understood in a sense defined in \eqref{distdet}.
In the following lemma, we obtain the (local) strong forms of Equations \eqref{CurlLambdaDistributional} and \eqref{CompatibiilityDistributional} as a result of specific regularity assumptions on the distributional strain fields.     
\begin{lemma}
\label{CompatibilityLemma}
For a simply connected open set $\Omega \subset \mathbb{R}^2$ consider $\boldsymbol{E}\in \mathcal{B}(\Omega,\Sym)$ satisfying $\deg(\boldsymbol{E})<0$, such that $\boldsymbol{E}(\boldsymbol{\phi})=\int_\Omega  \left\langle \boldsymbol{e}, \boldsymbol{\phi} \right\rangle \da$ for all $\boldsymbol{\phi} \in \mathcal{D}(\Omega-O,\Lin)$, and $\boldsymbol{\Lambda}\in \mathcal{D}'(\Omega,\Sym)$ satisfying $\deg(\boldsymbol{\Lambda})<0$, where $\boldsymbol{\Lambda}=\boldsymbol{\Lambda}_1 + \boldsymbol{\Lambda}_2$ with $\boldsymbol{\Lambda}_1\in \mathcal{B}(\Omega,\Sym)$ and $\boldsymbol{\Lambda}_1\in \mathcal{C}(\Omega,\Sym)$ such that $\boldsymbol{\Lambda}_1(\boldsymbol{\phi})=\int_\Omega  \left\langle \boldsymbol{\lambda}, \boldsymbol{\phi} \right\rangle \da$ and  $\boldsymbol{\Lambda}_2(\boldsymbol{\phi})=\int_S \left\langle \boldsymbol{\gamma}, \boldsymbol{\phi} \right\rangle \dl$ for all $\boldsymbol{\phi} \in \mathcal{D}(\Omega-O,\Lin)$. Then

\noindent (a) There exist $ {W}\in \mathcal{B}(\Omega)$ with bulk density $w$ such that $w$ is a piecewise smooth scalar field continuous across $S$, satisfying Equations \eqref{DefinitionCurvatureStrainBulk} and \eqref{DefinitionCurvatureStrainConcentration}, if and only if 
\begin{subequations}
\label{CurlLambdaCompatibility}
\begin{align}
\curl \boldsymbol{\lambda}=\boldsymbol{0} ~\text{in} ~\Omega-\{S\cup \{O\}\},
\\
\llbracket \boldsymbol{\lambda} \rrbracket \boldsymbol{t} + \frac{d }{ds} (\boldsymbol{\gamma} \boldsymbol{\nu})= \boldsymbol{0} ~\text{on}~ S-\{O\},
\\
 \boldsymbol{\gamma} \times \boldsymbol{\nu}= \boldsymbol{0} ~\text{on}~ S-\{O\},~\text{and}
\\
\int_{\partial B_\epsilon -S}  \boldsymbol{\lambda} \boldsymbol{t} \dl + \sum \boldsymbol{\gamma} (\partial B_\epsilon \cap S)\boldsymbol{\nu} = 0, 
\end{align}
\end{subequations} 
where the summation $\Sigma$ is over all points of intersection of the loop $\partial B_\epsilon$ with $S.$

\noindent (b) Assuming that $\boldsymbol{\Lambda}$ satisfy Equations \eqref{CurlLambdaCompatibility}, with a $W$ which satisfies \eqref{DefinitionCurvatureStrainBulk} and \eqref{DefinitionCurvatureStrainConcentration}, there exist $\boldsymbol{U} \in \mathcal{B}(\Omega,\mathbb{R}^2)$, with $\boldsymbol{U}(\boldsymbol{\psi})=\int_\Omega \left\langle \boldsymbol{u},\boldsymbol{\psi}\right\rangle \da$ for all $\boldsymbol{\psi} \in \mathcal{D}(\Omega,\mathbb{R}^2)$ and $\boldsymbol{u}$ is a piecewise smooth vector field continuous across $S$, satisfying \eqref{DefinitionStretchingStrainBulk} if and only if 
\begin{subequations}
\label{GaussiamCurvatureCompatibiltiy}
\begin{align}
\curl \curl \boldsymbol{e} + \det (\boldsymbol{\lambda})={0} ~\text{in} ~\Omega-\{S\cup \{O\}\},
\\
\left\langle \{\boldsymbol{\lambda} \}, \boldsymbol{t}\otimes \boldsymbol{t} \right\rangle \left\langle \boldsymbol{\gamma}, \boldsymbol{\nu}\otimes \boldsymbol{\nu} \right\rangle +\left\langle \left\llbracket \nabla \boldsymbol{e}\right\rrbracket, \boldsymbol{r}  \right\rangle + k \left\langle \left\llbracket \boldsymbol{e}\right\rrbracket, \boldsymbol{\nu}\otimes \boldsymbol{\nu}  \right\rangle = {0} ~\text{on}~ S-\{O\},
\\
\langle \llbracket \boldsymbol{e} \rrbracket, \boldsymbol{t}\otimes \boldsymbol{t} \rangle = {0} ~\text{on}~ S-\{O\},~\text{and}
\\
\int_{\partial B_\epsilon -S}  \big(\boldsymbol{\varepsilon}\boldsymbol{t} + \left( (\boldsymbol{x}-\boldsymbol{x}_0)\times\boldsymbol{e}_3\right) \langle \curl \boldsymbol{\varepsilon}, \boldsymbol{t} \rangle \big) \dl  
 + \sum \big( (\boldsymbol{x}-\boldsymbol{x}_0)\times \boldsymbol{e}_3 \big)\langle\boldsymbol\llbracket \boldsymbol{\varepsilon} \rrbracket (\partial B_\epsilon \cap S),\boldsymbol{t} \otimes \boldsymbol{\nu}\rangle =0,
\end{align}
\end{subequations} 
where $\boldsymbol{r}$ is as defined in \eqref{defrs}$_1$ and $\boldsymbol{\varepsilon} = \boldsymbol{e}-{\nabla w \otimes \nabla w}$. The summation $\Sigma$ is over all points of intersection of the loop $\partial B_\epsilon$ with $S.$
\begin{proof}
\noindent (a)  Lemma~\ref{CurlRestrictionLemma}, in conjunction with \cite[Id.~2.3]{pandey2020topological}, implies the equivalence of \eqref{CurlLambdaCompatibility} and $\Curl \boldsymbol{\Lambda}=\boldsymbol{0}$. The existence and the required regularity of $W$, which satisfies $\nabla\nabla W =\boldsymbol{\Lambda}$, follows from Lemma~\ref{SymFieldsAiryHessian}.

\noindent (b) Consider $\boldsymbol{T}\in \mathcal{B}(\Omega,\Sym)$ given by $\boldsymbol{T}(\boldsymbol{\psi})=\int_\Omega \langle \boldsymbol{\varepsilon},\boldsymbol{\psi}\rangle \da$ for all $\boldsymbol{\psi}\in \mathcal{D}(\Omega,\Lin).$ According to Identity~\ref{CurlCurlFormulas} and Lemma~\ref{CurlCurlRestrictionzeroLemma}, Equations \eqref{GaussiamCurvatureCompatibiltiy} are equivalent to $\Curl \Curl \boldsymbol{T}=0$. Thereupon Lemma~\ref{StrainCompatibilityB} implies the existence of $\boldsymbol{U}$, of the required regularity such that $({1}/{2})(\nabla \boldsymbol{U} +\nabla \boldsymbol{U}^T)=\boldsymbol{T}$, or a piecewise smooth vector field $\boldsymbol{u}$  continuous across $S$ satisfying \eqref{DefinitionStretchingStrainBulk}. 
\end{proof}
\end{lemma}
Equations \eqref{CurlLambdaCompatibility} and \eqref{GaussiamCurvatureCompatibiltiy} are necessary and sufficient for the existence of piecewise smooth fields $\boldsymbol{u}$, $w$, continuous across $S$, satisfying \eqref{DefinitionofStrains}. It is important to note that compatibility equations \eqref{GaussiamCurvatureCompatibiltiy} can be stated unambiguously only after the existence of the $w$ field is established as a consequence of \eqref{CurlLambdaCompatibility}.

\subsection{Singular sources of inhomogeneity}\label{sources}

In this section we generalize the incompatibility relations \eqref{vksmooth3} in terms of distributional strains and distributional sources of inhomogeneity. We postulate an additive decomposition of the distributional strain fields into elastic and plastic components; i.e., we decompose stretching strain $\boldsymbol{E}\in \mathcal{D}'(\Omega,\Sym)$ as $\boldsymbol{E}=\boldsymbol{E}^e + \boldsymbol{E}^p$, where $\boldsymbol{E}^e\in \mathcal{D}'(\Omega,\Sym)$ and $\boldsymbol{E}^p\in \mathcal{D}'(\Omega,\Sym)$ represent elastic and plastic stretching strain fields, respectively, and  bending strain $\boldsymbol{\Lambda}\in \mathcal{D}'(\Omega,\Sym)$ as $\boldsymbol{\Lambda}=\boldsymbol{\Lambda}^e + \boldsymbol{\Lambda}^p$, where $\boldsymbol{\Lambda}^e\in \mathcal{D}'(\Omega,\Sym)$ and $\boldsymbol{\Lambda}^p\in \mathcal{D}'(\Omega,\Sym)$ represent elastic and plastic bending strains, respectively. Unlike $\boldsymbol{E}$ and $\boldsymbol{\Lambda}$, the elastic and plastic strain fields are not compatible, i.e., they do not satisfy conditions of the type \eqref{CurlLambdaDistributional} and \eqref{CompatibiilityDistributional}. In other words, there exist non-trivial incompatibility fields $N_1 \in \mathcal{D}'(\Omega)$ and $\boldsymbol{N}_2 \in \mathcal{D}'(\Omega,\mathbb{R}^2)$ such that $N_1=\Curl \Curl \boldsymbol{E}^p+\Det(\boldsymbol{\Lambda}^p)$ and $\boldsymbol{N}_2=\Curl \boldsymbol{\Lambda}^p$; the plastic strain fields are compatible if and only if $N_1=0$ and $\boldsymbol{N}_2=\boldsymbol{0}$. Clearly, $N_1$ is well defined only if $\Det(\boldsymbol{\Lambda}^p)$ is well defined, which in turn requires us to impose some further regularity assumptions on $\boldsymbol{\Lambda}^p$ (as discussed in Section~\ref{mabrack}). However, depending on how we prescribe the inhomogeneous terms, $N_1$ may or may not appear in our governing equations. There are two choices, either $\boldsymbol{E}^p$ and $\boldsymbol{\Lambda}^p$ or $N_1$ and $\boldsymbol{\Lambda}^p$ are prescribed as inhomogeneity sources. In order to elaborate, we discuss these two cases separately:
\begin{enumerate}
\item \textit{$\boldsymbol{E}^p$ and $\boldsymbol{\Lambda}^p$ prescribed}: We assume that $\boldsymbol{E}^p\in \mathcal{D}'(\Omega,\Sym)$ and $\boldsymbol{\Lambda}^p\in \mathcal{D}'(\Omega,\Sym)$ are given. This would be so whenever we are in a situation to prescribe the plastic strains directly, e.g., in classical problems of plastic deformation and the problems of growth. We do not have to then work with incompatibility fields. The compatibility condition~\eqref{CompatibiilityDistributional}, on using the additive decomposition for $\boldsymbol{E}$, yields
\begin{equation}
\label{N1IncompatibilityElasticStrain}
\Curl\Curl \boldsymbol{E}^e + \Det(\boldsymbol{\Lambda})=-\Curl \Curl \boldsymbol{E}^p.
\end{equation} \label{case1}%
\item  \textit{$N_1$ and $\boldsymbol{\Lambda}^p$ prescribed}: We assume that $N_1 \in \mathcal{D}'(\Omega)$ and $\boldsymbol{\Lambda}^p\in \mathcal{D}'(\Omega,\Sym)$ are given. This would be the situation when we describe the inhomogeneity explicitly in terms of defect densities. The incompatibility $N_1$ is directly related to a distribution of dislocations, disclinations, and metric anomalies, see Remark~\ref{defects} below. Equation~\eqref{N1IncompatibilityElasticStrain} can be rewritten with sources in terms of $N_1$ and $\boldsymbol{\Lambda}^p$ as
\begin{equation}
\Curl\Curl \boldsymbol{E}^e + \Det(\boldsymbol{\Lambda})=-N_1+ \Det(\boldsymbol{\Lambda}^p). \label{N1IncompatibilityElasticStrain1}%
\end{equation}
The determinant of $\boldsymbol{\Lambda}^p$ in \eqref{N1IncompatibilityElasticStrain1} is however meaningful only if further regularity restrictions are imposed on $\boldsymbol{\Lambda}^p$. 
We note that, for any open set $\omega\subset \Omega$, if $\boldsymbol{\Lambda}^p|_{\omega}$ is a continuous field then $\Det(\boldsymbol{\Lambda}^p)$ can be interpreted in terms of the usual determinant $\det(\boldsymbol{\Lambda}^p|_{\omega})$. This would however preclude us from considering plastic bending strain fields which are discontinuous at $S$, or are unbounded in $\omega$, or, most importantly, those which concentrate on $S$. Such $\boldsymbol{\Lambda}^p$ can be only considered in $\omega$ only when $\boldsymbol{N}_2|_{\omega}=\boldsymbol{0}$. The latter allows us to have further regularity assumptions on $\boldsymbol{\Lambda}^p$ to make $\Det(\boldsymbol{\Lambda}^p)$ meaningful. Therefore, \eqref{N1IncompatibilityElasticStrain1} is well defined in any local neighborhood of $\Omega$ wherein either $\boldsymbol{\Lambda}^p$ is continuous or $\boldsymbol{N}_2$ vanishes identically. There are no such restrictions on $\boldsymbol{\Lambda}^p$ in Case~\ref{case1}.  \label{case2}
\end{enumerate}
Equation \eqref{N1IncompatibilityElasticStrain} or \eqref{N1IncompatibilityElasticStrain1} (depending on the nature of prescribed fields) will form the basis of deriving one of the generalized von K{\'a}rm{\'a}n equations.

\begin{remark}\label{defects}(Incompatibility in terms of defect fields)
 Given a distributional disclination density $\boldsymbol{\Theta}\in \mathcal{D}'(\Omega),$ a dislocation density $\boldsymbol{A} \in \mathcal{D}'(\Omega,\mathbb{R}^2)$, and in-plane metric anomaly strain field $\boldsymbol{Q}\in \mathcal{D}'(\Omega,\Sym)$, we postulate the relationship of defect fields with the strain incompatibility field as
\begin{equation}
\label{IncompatibilityEquation1}
{\Theta} + \Curl \boldsymbol{A} + \Curl \Curl \boldsymbol{Q}=N_1.
\end{equation}
This relation allows us to incorporate strain incompatibility arising from bulk and interfacial defect fields as well as those supported at $O$.  We make further regularity assumptions on the defect fields.
We consider $\Theta(\psi)=\int_\Omega \theta_B \psi \da+\int_S \theta_S \psi \dl$, for all $\psi \in \mathcal{D}(\Omega-O)$, where $ \theta_B$ and $\theta_S$ are bulk and interfacial densities of disclinations, respectively; $\boldsymbol{A}(\boldsymbol{\psi})=\int_\Omega \langle\boldsymbol{\alpha}_B, \boldsymbol{\psi}\rangle \da+\int_S \langle\boldsymbol{\alpha}_S, \boldsymbol{\psi}\rangle \dl$, for all $\boldsymbol{\psi} \in \mathcal{D}(\Omega-O,\mathbb{R}^2)$, where $ \alpha_B$ and $\alpha_S$ are bulk and interfacial densities of dislocations, respectively; and $\boldsymbol{Q}(\boldsymbol{\psi})=\int_\Omega \langle \boldsymbol{q},\boldsymbol{\psi} \rangle \da$, for all $\boldsymbol{\psi}\in \mathcal{D}(\Omega-O,\Lin)$, where $\boldsymbol{q}$ is the bulk density of metric anomalies. On substituting these into \eqref{IncompatibilityEquation1}, and using \cite[Id. 2.3]{pandey2020topological} and Identity~\ref{CurlCurlFormulas}, we obtain incompatibility field of the form $N_1(\psi)=\int_\Omega {\eta} \psi \da+\int_S {\zeta}_{1} \psi \dl+\int_S {\zeta}_{2} \partial \psi/\partial \boldsymbol{\nu} \dl$, for all $\psi \in \mathcal{D}(\Omega-O)$, such that 
\begin{subequations}
\label{N1Restriction}
\begin{align}
\curl \boldsymbol{\alpha}_B + \theta_B + \curl \curl \boldsymbol{q}= {\eta}~ \text{in}~ \Omega-\{S\cup\{O\} \}, \label{N1Restriction1}
\\
 \frac{d}{d s}\langle \llbracket \boldsymbol{\alpha}_B \rrbracket,\boldsymbol{t} \rangle+ \langle  \boldsymbol{\alpha}_{S},\boldsymbol{\nu}  \rangle + \theta_S + \left\langle \left\llbracket \nabla \boldsymbol{q}\right\rrbracket, \boldsymbol{r} \right\rangle + k \left\langle \left\llbracket \boldsymbol{q}\right\rrbracket, \boldsymbol{\nu}\otimes \boldsymbol{\nu}  \right\rangle= {\zeta_1} ~\text{on}~ S-\{O\}, \label{N1Restriction2}
\\
\langle  \boldsymbol{\alpha}_{S},\boldsymbol{t}  \rangle + \left\langle \left\llbracket \boldsymbol{q}\right\rrbracket, \boldsymbol{t}\otimes \boldsymbol{t}  \right\rangle = {\zeta_2}~ \text{on}~ S-\{O\}, and \label{N1Restriction3}
\\
\Theta(v^\alpha)+\boldsymbol{A}(\boldsymbol{e}_3 \times \nabla v^\alpha)+\boldsymbol{Q}(\mathbb{A}\nabla\nabla v^\alpha)=N_1(v^\alpha)~\text{at}~O, \label{N1Restriction4}
\end{align}
\end{subequations}
for all multi indices $\alpha \in \mathbb{N}^2$ satisfying $|\alpha| \leq q$, where $q$ is the maximum of $\deg(\Theta)$, $\deg(\boldsymbol{A})+1$, and $\deg(\boldsymbol{Q})+2$, and $v^\alpha$ are as defined in Lemma~\ref{RepresentationELemma}. In addition, $\deg(N_1)\leq q$. Equations \eqref{N1Restriction} are in fact equivalent to \eqref{IncompatibilityEquation1}. Indeed, \eqref{N1Restriction1}-\eqref{N1Restriction3} imply that $({\Theta} + \Curl \boldsymbol{A} + \Curl \Curl \boldsymbol{Q}-N_1)\in \mathcal{E}(\Omega)$. Equation \eqref{N1Restriction4}, on using Lemma~\ref{RepresentationELemma}, then implies \eqref{IncompatibilityEquation1}.
\end{remark}

\subsection{Generalized equilibrium conditions} \label{geneqb}
In this section we introduce a distributional form of the equilibrium conditions (thereby generalizing \eqref{vksmooth4}). The stress and moment fields are accordingly generalized as distributions. In our general setting we allow for stress and moments to develop concentrations on $S$. We obtain strong form implications of the distributional conditions pointwise in $\Omega$, away from $S$ and $O$, on $S$ and at $O$. Consider a distributional stress field $\boldsymbol{\Sigma}\in \mathcal{D}'(\Omega,\Sym)$ which satisfies 
\begin{equation}
\Div \boldsymbol{\Sigma}=\boldsymbol{0}. \label{EquilibriumInPlaneStressDistributional}
\end{equation}
For a simply connected $\Omega$, Lemma~\ref{SymFieldsAiryHessian} implies the existence of $\Phi \in \mathcal{D}'(\Omega)$ such that $\boldsymbol{\Sigma}=\mathbb{A}\nabla \nabla \Phi$. Consider a distributional moment field $\boldsymbol{M}\in \mathcal{D}'(\Omega,\Sym)$ and a bending strain field $\boldsymbol{\Lambda}\in \mathcal{D}'(\Omega,\Sym)$ such that $\Curl \boldsymbol{\Lambda}=\boldsymbol{0}$. Therefore there exist $W\in \mathcal{D}'(\Omega)$ which satisfies $\boldsymbol{\Lambda}=\nabla \nabla W$. We require $\Phi$ and $W$ to be fields which satisfy the assumptions mentioned in either \ref{a1} or \ref{a2}. Consequently an inner product of $\boldsymbol{\Sigma}$ and $\boldsymbol{\Lambda}$ can be introduced in a distributional sense (as in \eqref{innerprod}). We also consider a body force field in terms of distributional transverse force field $F \in \mathcal{D}'(\Omega)$.
The second equilibrium condition can then be postulated in the form
\begin{equation}
\label{EquilibriumMomentDistributional1}
\Div \Div \boldsymbol{M} - \langle\boldsymbol{\Sigma},\boldsymbol{\Lambda} \rangle=F
\end{equation}
which can be equivalently written as
\begin{equation}
\Div\Div \boldsymbol{M} - [\Phi,W]=F, \label{EquilibriumMomentDistributional2}
\end{equation}
where the distributional Monge-Amp{\`e}re bracket is as introduced in \ref{a1} or \ref{a2} (depending on the regularity of $\Phi$ and $W$). In the following we will obtain the local form of the distributional equilibrium equations~\eqref{EquilibriumInPlaneStressDistributional} and \eqref{EquilibriumMomentDistributional1} under further regularity assumptions on $\boldsymbol{\Sigma}, \boldsymbol{M}, \boldsymbol{\Lambda}$, and $F$.

We consider a stress field such that $\boldsymbol{\Sigma}=\boldsymbol{\Sigma}_1+\boldsymbol{\Sigma}_2$ where $\boldsymbol{\Sigma}_1\in \mathcal{B}(\Omega,\Sym)$, with bulk density $\boldsymbol{\sigma}$, and $\boldsymbol{\Sigma}_2 \in \mathcal{C}(\Omega,\Sym)$, with line density $\boldsymbol{\tau}$; a moment field such that $\boldsymbol{M}=\boldsymbol{M}_1+\boldsymbol{M}_2$ where $\boldsymbol{M}_1\in \mathcal{B}(\Omega,\Sym)$, with bulk density $\boldsymbol{m}$, and $\boldsymbol{M}_2 \in \mathcal{C}(\Omega,\Sym)$, with line density $\boldsymbol{n}$; a bending strain field such that $\boldsymbol{\Lambda}=\boldsymbol{\Lambda}_1+\boldsymbol{\Lambda}_2$ where $\boldsymbol{\Lambda}_1\in \mathcal{B}(\Omega,\Sym)$, with bulk density $\boldsymbol{\lambda}$, and $\boldsymbol{\Lambda}_2 \in \mathcal{C}(\Omega,\Sym)$, with line density $\boldsymbol{\gamma}$; and a transverse force field such that $F=F_1+F_2$ where $F_1\in \mathcal{B}(\Omega)$, with bulk density $f_1$, and $F_2 \in \mathcal{C}(\Omega)$, with line density $f_2$. We can then use the divergence identities from \cite[Id. 2.2]{pandey2020topological} to derive the local form of \eqref{EquilibriumInPlaneStressDistributional}:
\begin{subequations}
\label{EquilibriumEquationsDiscontinuousStressFields}
\begin{align}
\div \boldsymbol{\sigma}= \boldsymbol{0} ~\text{in}~\Omega-\{S\cup \{O\} \}, \label{EquilibriumStress1}
\\
 \frac{d\boldsymbol{\tau}}{ds} \boldsymbol{t} - \llbracket \boldsymbol{\sigma} \rrbracket \boldsymbol{\nu}=\boldsymbol{0}  ~\text{on}~S-\{O\}, \label{EquilibriumStress2}
\\
\boldsymbol{\tau} \boldsymbol{\nu}=\boldsymbol{0}  ~\text{on}~S-\{O\}, ~\text{and} \label{EquilibriumStress3}
\\
\label{EquilibriumStressSingular}
\boldsymbol{\Sigma}(\nabla (v^\alpha \boldsymbol{e}_1))=0, ~~
\boldsymbol{\Sigma}(\nabla (v^\alpha \boldsymbol{e}_2))=0~\text{at}~O,
\end{align}
\end{subequations}
for all multi indices $\alpha \in \mathbb{N}^2$ satisfying $|\alpha| \leq \deg(\boldsymbol{\Sigma})+1$, where $v^\alpha$ are as defined in Lemma~\ref{RepresentationELemma}. The local equations  \eqref{EquilibriumEquationsDiscontinuousStressFields} are equivalent to \eqref{EquilibriumInPlaneStressDistributional}. Indeed, \eqref{EquilibriumStress1}-\eqref{EquilibriumStress3} imply $(\Div \boldsymbol{\Sigma})|_{\Omega-O}=\boldsymbol{0}$ or, in other words, $\Div \boldsymbol{\Sigma} \in \mathcal{E}(\Omega,\mathbb{R}^2)$. Equation \eqref{EquilibriumStressSingular}, on using Lemma~\ref{RepresentationELemma}, then implies \eqref{EquilibriumInPlaneStressDistributional}. Similarly, we use the divergence identities \cite[Id. 2.2]{pandey2020topological}, in addition to Identity~\ref{CurlCurlFormulas}, to deduce the local form of \eqref{EquilibriumMomentDistributional1}:
\begin{subequations}
\label{MomentBalanceInterface}
\begin{align}
\div\div \boldsymbol{m} - \langle \boldsymbol{\sigma}, \boldsymbol{\lambda} \rangle=f_1 ~\text{in}~\Omega-\{S\cup \{O\}\}, \label{MomentBalance1}
\\
\begin{aligned}
\left\langle \llbracket \div \boldsymbol{m} \rrbracket, \boldsymbol{\nu} \right\rangle+\left\langle \frac{d\llbracket\boldsymbol{m}\rrbracket}{ds} ,\boldsymbol{t}\otimes \boldsymbol{\nu} \right\rangle -\left\langle \frac{d^2\boldsymbol{n}}{ds^2},\boldsymbol{t}\otimes \boldsymbol{t} \right\rangle - k^2\langle\boldsymbol{n},\boldsymbol{t}\otimes \boldsymbol{t} \rangle-k\langle\llbracket\boldsymbol{m}\rrbracket,\boldsymbol{t}\otimes \boldsymbol{t} \rangle  \\ + \left\langle \{\boldsymbol{\sigma}\}, \boldsymbol{\nu}\otimes \boldsymbol{\nu} \right\rangle \left\langle \boldsymbol{\gamma}, \boldsymbol{\nu}\otimes \boldsymbol{\nu} \right\rangle + \left\langle \{\boldsymbol{\lambda}\}, \boldsymbol{t}\otimes \boldsymbol{t} \right\rangle \left\langle \boldsymbol{\tau}, \boldsymbol{t}\otimes \boldsymbol{t} \right\rangle =f_2 ~\text{on}~S-\{O\}, 
\end{aligned} \label{MomentBalance2}
\\
\left\langle\frac{d\boldsymbol{n}}{ds},\boldsymbol{t}\otimes\boldsymbol{\nu}\right\rangle - k \left\langle \boldsymbol{n},\boldsymbol{t}\otimes \boldsymbol{t} \right\rangle - \langle \llbracket \boldsymbol{m}\rrbracket,\boldsymbol{\nu}\otimes \boldsymbol{\nu} \rangle =0 ~\text{on}~S-\{O\}, \label{MomentBalance3}
\\
\left\langle   \boldsymbol{n} , \boldsymbol{\nu}\otimes\boldsymbol{\nu} \right\rangle =0 ~\text{on}~S-\{O\},~\text{and} \label{MomentBalance4}
\\
\label{MomentBalanceSingularPoint}
\boldsymbol{M}(\nabla\nabla v^\alpha)+\nabla \Phi \otimes \nabla W (\mathbb{A} \nabla\nabla v^\alpha)=F(v^\alpha)~\text{at}~O,
\end{align}
\end{subequations}
for all multi indices $\alpha \in \mathbb{N}^2$ satisfying $|\alpha|\leq q$, where $q$ is the maximum of $\deg(\boldsymbol{M})+2$, $\deg(\nabla \Phi \otimes \nabla W)+2$, and $\deg(F)$. 
The local equations Equations \eqref{MomentBalanceInterface} are equivalent to \eqref{EquilibriumMomentDistributional1}. Indeed, as a consequence of \eqref{MomentBalance1}-\eqref{MomentBalance4}, $\Div \Div \boldsymbol{M} + \langle\boldsymbol{\Sigma},\boldsymbol{\Lambda} \rangle - F \in \mathcal{E}(\Omega).$ Equation \eqref{MomentBalanceSingularPoint}, on using Lemma~\ref{RepresentationELemma}, then implies 
\eqref{EquilibriumMomentDistributional1}.

\begin{remark} (Equilibrium equation with point supported force field) We consider a situation where $S = \emptyset$ and the singular support of the fields is a subset of $\{O\}$. Let $\deg(\boldsymbol{M}) < 0$. The transverse force field consists of an isolated point force and a point dipole both acting at $O$, i.e.,  $F={f}_0 \delta_O+\langle {\boldsymbol{f}}_1,\nabla \delta_O\rangle$, where ${f}_0$ represents the magnitude of the point force and ${\boldsymbol{f}}_1$ is the force dipole vector. Using Lemma~\ref{DivDivLemma} we can establish the equivalence of the equilibrium condition~\eqref{EquilibriumMomentDistributional2} with the local equations
\begin{subequations}
\begin{align}
\label{EquilibriumBulkk}
\div\div \boldsymbol{m}-[\phi,w]=0~\text{in}~\Omega-\{O\}~\text{and}
\\
\label{LoopIntegralPoinForceDipole}
\int_{\partial B_\epsilon} (\boldsymbol{a}\boldsymbol{\nu}-(\boldsymbol{x}-\boldsymbol{x}_0)\langle\div \boldsymbol{a},\boldsymbol{\nu} \rangle) \dl=-{f}_0\boldsymbol{x}_0 +{\boldsymbol{f}}_1,
\end{align}
\end{subequations}
for all $\boldsymbol{x}_0 \in \mathbb{R}^2$, where $\boldsymbol{a}=\boldsymbol{m}-(\boldsymbol{e}_3\times \nabla \phi) \otimes (\boldsymbol{e}_3\times \nabla w)$. Note that \eqref{LoopIntegralPoinForceDipole} is equivalent to \eqref{MomentBalanceSingularPoint} under the present assumptions.
\end{remark}

\begin{remark}(Equilibrium condition as a sequential limit of smooth fields) Consider a stress field such that $\boldsymbol{\Sigma} \in \mathcal{B}(\Omega,\Sym)$, with $\boldsymbol{\sigma}$ bounded in $\Omega$, and a bending strain field (as considered above) with $\boldsymbol{\lambda}$ and $\boldsymbol{\gamma}$ bounded in $\Omega$ and $S$, respectively. We let the moment field to be a general distribution $\boldsymbol{M} \in \mathcal{D}'(\Omega,\Sym)$. We also assume $F=0$. Then, according to Lemma \ref{SymFieldsAiryHessian}, there exist $\Phi \in \mathcal{B}(\Omega)$, such that $\nabla \Phi$ is continuous across $S$ and $\boldsymbol{\Sigma}=\mathbb{A} \nabla \nabla \Phi$, and $W\in \mathcal{B}(\Omega)$ continuous across $S$, such that $\boldsymbol{\Lambda}=\nabla\nabla W$. The non-smooth fields $W,\Phi$, and $\boldsymbol{M}$ can be seen as the weak limits of sequences of smooth fields $w_n$, $\phi_n$, and $\boldsymbol{m}_n$ such that $\nabla w_n$ and $\nabla \nabla \phi_n$ converge to $\nabla W$ and $\nabla \Phi$ weakly in $L^2$ and $\boldsymbol{m}_n$ converges to $\boldsymbol{M}$ in the distributional sense~\cite{evans1990weak}. The stress field $\boldsymbol{\Sigma}$ is the $L^2$ limit of the sequence of smooth symmetric tensor fields $\boldsymbol{\sigma}_n=\mathbb{A}\nabla \nabla \phi_n.$ The equilibrium condition, in terms of the smooth fields $\boldsymbol{m}_n$, $\boldsymbol{\sigma}_n$, and $w_n$, is $\div\div \boldsymbol{m}_n - [\phi_n,w_n]=0 ~\text{in}~\Omega$. 
Recall that $[\phi_n,w_n]$ converges to $[\Phi,W]$ and $\div \div \boldsymbol{m}_n$ converges to $\Div\Div \boldsymbol{M}$ in the sense of distributions, see Remark~\ref{maseqlim}. Therefore for weak fields, with the given regularity (bounded densities of stress and bending strain fields and no concentration in stress), equilibrium condition \eqref{EquilibriumMomentDistributional1} can be interpreted as the weak sequential limit of a sequence of equilibrium conditions with smooth fields. 
\end{remark}
 
\subsection{Generalized von K{\'a}rm{\'a}n equations} \label{genvkequations}
We now derive the generalized von K{\'a}rm{\'a}n equations both in their distributional and local forms. The latter includes pointwise equations in the domain $\Omega$, away from $S$ and $O$, on the interfacial curve S, away from $O$ (i.e., if $O$ lies on $S$), and at $O$. We assume that both the stress field and the moment field does not concentrate on $S$ (this is less general than what was considered in writing the equilibrium equations); i.e., $\boldsymbol{\Sigma}\in \mathcal{B}(\Omega,\Sym)$, with piecewise smooth bulk density $\boldsymbol{\sigma}$, and $\boldsymbol{M} \in \mathcal{B}(\Omega,\Sym)$, with piecewise smooth bulk density $\boldsymbol{m}$. We also assume the elastic strain fields to not concentrate on $S$, i.e., $\boldsymbol{E}^e\in \mathcal{B}(\Omega,\Sym)$ and $\boldsymbol{\Lambda}^e \in \mathcal{B}(\Omega,\Sym)$, with bulk densities $\boldsymbol{e}^e$ and $\boldsymbol{\lambda}^e$, respectively. We assume constitutive relations as given in~\eqref{vksmooth5}, which can be equivalently written as
\begin{equation}
\label{IsotropicConstitutiveRelations}
\boldsymbol{\Sigma}=\frac{E}{1-\nu^2}\left( (1-\nu)\boldsymbol{E}^e+ \nu \tr(\boldsymbol{E}^e) \boldsymbol{I} \right)~\text{and} ~
\boldsymbol{M}=D\left( (1-\nu)\boldsymbol{\Lambda}^e+ \nu \tr(\boldsymbol{\Lambda}^e) \boldsymbol{I} \right).
\end{equation}
The stress field satisfies the equilibrium equation \eqref{EquilibriumInPlaneStressDistributional} which, for the given regularity of $\boldsymbol{\Sigma}$, is equivalent to the existence of a stress function $\Phi \in \mathcal{B}(\Omega)$, with bulk density $\phi$, such that $\boldsymbol{\Sigma}=\mathbb{A}\nabla\nabla \Phi$, see Lemma~\ref{SymFieldsAiryHessian}. The bending strain is such that $\boldsymbol{\Lambda}=\boldsymbol{\Lambda}_1+\boldsymbol{\Lambda}_2$, where $\boldsymbol{\Lambda}_1\in \mathcal{B}(\Omega,\Sym)$ and $\boldsymbol{\Lambda}_2 \in \mathcal{C}(\Omega,\Sym)$. It satisfies the compatibility equation~\eqref{CurlLambdaDistributional}. According to Lemma~\ref{SymFieldsAiryHessian} there exist $W\in \mathcal{B}(\Omega)$, with bulk density $w$, such that $\boldsymbol{\Lambda}=\nabla\nabla W$. Both $w$ and $\phi$ are continuous across $S$. We note that $\nabla w$ can be discontinuous across  $S$, since $\boldsymbol{\Lambda}$ concentrates on the interface, whereas $\llbracket \nabla \phi \rrbracket=\boldsymbol{0}$ since we have assumed the stress to not concentrate on $S$. Furthermore, we assume $W$ to satisfy the assumptions in \ref{a2p} and $\Phi$ and $W$ to jointly satisfy the assumptions in \ref{a2}. The former allows us to have a well defined generalized determinant $\Det(\boldsymbol{\Lambda})$ while the latter allows us to have a well defined generalized inner product $\langle\boldsymbol{\Sigma},\boldsymbol{\Lambda}\rangle$.

In order to derive the von K{\'a}rm{\'a}n equations we need to choose between the two options (as given in Section~\ref{sources}) for the specification of inhomogeneity. The regularity of the considered ${\boldsymbol{\Lambda}^p}$ depends on the choice. Accordingly, we will present the derivation for the two cases separately.

\noindent \textit {Case 1. $\boldsymbol{E}^p$ and ${\boldsymbol{\Lambda}^p}$ prescribed}. We consider plastic strains such that  $\boldsymbol{E}^p \in \mathcal{B}(\Omega,\Sym)$, with bulk density $\boldsymbol{e}^p$, and $\boldsymbol{\Lambda}^p=\boldsymbol{\Lambda}^p_1+\boldsymbol{\Lambda}^p_2$, where $\boldsymbol{\Lambda}^p_1\in \mathcal{B}(\Omega,\Sym)$ and $\boldsymbol{\Lambda}^p_2 \in \mathcal{C}(\Omega,\Sym)$, with densities $\boldsymbol{\lambda}^p$ and $\boldsymbol{\gamma}^p$, respectively (i.e., $\boldsymbol{\gamma}^p$ is the concentration of plastic bending strain on $S$). Due to our assumption of vanishing concentration of $\boldsymbol{\Lambda}^e$, we have $\boldsymbol{\gamma}  = \boldsymbol{\gamma}^p$, i.e.,
\begin{equation}
 - \llbracket \nabla w \rrbracket \otimes \boldsymbol{\nu} = \boldsymbol{\gamma}^p. \label{plastbendstrconc}
\end{equation} 
The first von K{\'a}rm{\'a}n equation (in a distributional form) follows from \eqref{N1IncompatibilityElasticStrain}, after writing $\boldsymbol{E}^e$ in terms of $\boldsymbol{\Sigma}$ and then $\Phi$, and substituting ${\boldsymbol{\Lambda}}$ in terms of $W$, as 
\begin{equation}
\frac{1}{E}\Delta^2 \Phi +\frac{1}{2} [W,W]=-\Curl\Curl \boldsymbol{E}^p. \label{vk1dist1}
\end{equation}
 The equivalent local form of  \eqref{vk1dist1} can be obtained using Identities~\ref{CurlCurlFormulas} and \ref{LaplacianFormula} as
\begin{subequations}
\label{vk1loc}
\begin{align}
\frac{1}{E}\Delta^2 \phi +\frac{1}{2} [w,w]=-\curl\curl \boldsymbol{e}^p ~ \text{in}~\Omega-\{S\cup\{O\}\}, \label{vk1loc1}
\\
\begin{aligned}
\frac{1}{E}\left\langle \llbracket \div \nabla\nabla \phi \rrbracket, \boldsymbol{\nu} \right\rangle + \frac{1}{E}\left\langle \frac{d}{ds} \llbracket\nabla\nabla \phi\rrbracket,\boldsymbol{t}\otimes \boldsymbol{\nu} \right\rangle + \left\langle \{ \nabla\nabla w \}, \boldsymbol{t}\otimes \boldsymbol{t} \right\rangle \left\langle \llbracket\nabla w \rrbracket, \boldsymbol{\nu} \right\rangle= 
\\
 -\left\langle \left\llbracket \nabla \boldsymbol{e}^p\right\rrbracket, \boldsymbol{r} \right\rangle- k \left\langle \left\llbracket \boldsymbol{e}^p\right\rrbracket, \boldsymbol{\nu}\otimes \boldsymbol{\nu}  \right\rangle ~ \text{on}~S-\{O\}, 
 \end{aligned} \label{vk1loc2}
\\
\frac{1}{E}\left\langle \llbracket\nabla\nabla \phi\rrbracket, \boldsymbol{\nu}\otimes \boldsymbol{\nu} \right\rangle=-\left\langle \llbracket\boldsymbol{e}^p\rrbracket ,\boldsymbol{t}\otimes\boldsymbol{t} \right\rangle ~\text{on} ~S-\{O\},~\text{and} \label{vk1loc3}
\\
\frac{1}{E}\Phi(\Delta^2 v^\alpha)-\nabla W\otimes \nabla W(\mathbb{A}\nabla\nabla v^\alpha)=-\boldsymbol{E}^p(\mathbb{A}\nabla\nabla v^\alpha) ~\text{at}~O, \label{vk1loc4}
\end{align}
\end{subequations}
for all multi indices $\alpha \in \mathbb{N}^2$ such that $|\alpha| \leq q$, where $q$ is the maximum of $\deg(\Phi)+4$, $\deg(\nabla W\otimes \nabla W)+2$, and $\deg(\boldsymbol{E}^p)+2$. Equations \eqref{vk1loc} are equivalent to \eqref{vk1dist1}. Indeed, \eqref{vk1loc1}-\eqref{vk1loc3} imply that $\Curl\Curl \boldsymbol{E}^e + \det(\boldsymbol{\Lambda})+\Curl \Curl \boldsymbol{E}^p \in \mathcal{E}(\Omega)$. Equation \eqref{vk1dist1} then follows  from \eqref{vk1loc4} using Lemma~\ref{RepresentationELemma}. The second von K{\'a}rm{\'a}n equation (in a distributional form) is obtained from \eqref{EquilibriumMomentDistributional2}, after writing $\boldsymbol{M}$ in terms of $\boldsymbol{\Lambda}^e$ (using \eqref{IsotropicConstitutiveRelations}$_2$) and subsequently in terms of $\boldsymbol{\Lambda}$ (hence $W$) and $\boldsymbol{\Lambda}^p$, as
\begin{equation}
D\Delta^2 W - [\Phi,W]=F+ D\left( (1-\nu)\Div \Div \boldsymbol{\Lambda}^p +\nu \Delta (\tr(\boldsymbol{\Lambda}^p))\right). \label{vk2dist1}
\end{equation}
 The equivalent local form of  \eqref{vk2dist1} can be obtained using \cite[Id. 2.2]{pandey2020topological}, in addition to Identities~\ref{CurlCurlFormulas} and \ref{LaplacianFormula}, as
\begin{subequations}
\label{VonKArmanEquilibriumStrongForms}
\begin{align}
D\Delta^2 w - [\phi,w]=f_1+ D\left( (1-\nu)\div \div {\boldsymbol{\lambda}^p} +\nu \Delta (\tr({\boldsymbol{\lambda}^p}))\right) ~ \text{in}~\Omega-\{S\cup\{O\}\},
\\
\begin{aligned}
D(1-\nu)\frac{d^2}{ds^2} \left\langle \llbracket \nabla w\rrbracket, \boldsymbol{\nu} \right\rangle + D\left\langle \llbracket \div \nabla\nabla w \rrbracket, \boldsymbol{\nu} \right\rangle - \left\langle \nabla\nabla \phi, \boldsymbol{t}\otimes \boldsymbol{t} \right\rangle \left\langle \llbracket\nabla w\rrbracket, \boldsymbol{\nu} \right\rangle = f_2 + \\D(1-\nu)\left(\left\langle \llbracket \div \boldsymbol{\lambda}^p \rrbracket, \boldsymbol{\nu} \right\rangle + \left\langle \frac{d}{ds} \llbracket\boldsymbol{\lambda}^p\rrbracket,\boldsymbol{t}\otimes \boldsymbol{\nu} \right\rangle +  k\langle\llbracket\boldsymbol{\lambda}^p\rrbracket,\boldsymbol{s} \rangle \right) + D\nu\llbracket \nabla(\tr \boldsymbol{\lambda}^p) \rrbracket   ~\text{on} ~S-\{O\},
\end{aligned}
\\
(1-\nu)k \left\langle \llbracket \nabla w\rrbracket, \boldsymbol{\nu} \right\rangle + \llbracket\Delta w\rrbracket = (1-\nu) \langle \llbracket\boldsymbol{\lambda}^p\rrbracket,\boldsymbol{\nu}\otimes \boldsymbol{\nu} \rangle+\nu \llbracket\tr(\boldsymbol{\lambda}^p)\rrbracket  ~\text{on} ~S-\{O\},~\text{and}
\\
D W(\Delta^2 v^\alpha)+\nabla \Phi \otimes \nabla W(\mathbb{A}\nabla\nabla v^\alpha)=F(v^\alpha) -D\left( (1-\nu)\boldsymbol{\Lambda}^p(\nabla\nabla v^\alpha) +\nu  (\tr(\boldsymbol{\Lambda}^p)(\Delta v^\alpha))\right)~\text{at}~O,
\end{align}
\end{subequations}
for all multi indices $\alpha \in \mathbb{N}^2$ such that $|\alpha| \leq q$, where $q$ is the maximum of $\deg(W)+4,$ $\deg(\nabla \Phi\otimes \nabla W)+2$, $\deg(F)$, and $\deg(\boldsymbol{\Lambda}^p)+2$. Equations \eqref{VonKArmanEquilibriumStrongForms} and \eqref{vk2dist1} are both equivalent (established using Lemma~\ref{RepresentationELemma}). We also note that  \eqref{plastbendstrconc} was used in deriving both \eqref{vk1loc} and \eqref{VonKArmanEquilibriumStrongForms}. The complete set of the desired von K{\'a}rm{\'a}n equations (in Case 1) is given as distributional equations in   \eqref{vk1dist1} and \eqref{vk2dist1} or, equivalently, as local pointwise equations in \eqref{plastbendstrconc}, \eqref{vk1loc}, and \eqref{VonKArmanEquilibriumStrongForms}.

\noindent \textit {Case 2. $N_1$ and ${\boldsymbol{\Lambda}^p}$ prescribed}.  We consider incompatibility  $N_1 \in \mathcal{D}'(\Omega)$ such that 
\begin{equation}
N_1(\psi)=\int_\Omega {\eta} \psi \da+\int_S {\zeta}_{1} \psi \dl+\int_S {\zeta}_{2} \frac{\partial \psi}{\partial \nu} \dl,
\end{equation}
for all $\psi \in \mathcal{D}(\Omega-O)$. The incompatibility field is related to the bulk and singular defect densities, as described pointwise in \eqref{N1Restriction}. We consider a continuous plastic bending strain ${\boldsymbol{\Lambda}^p}$. The bulk density $\boldsymbol{\lambda}^p$ is therefore continuous across $S$ and there are no concentrations in plastic strains. The continuity of $w$, combined with \eqref{plastbendstrconc} (with $\boldsymbol{\gamma}^p = \boldsymbol{0}$), then implies the continuity of $\nabla w$ across $S$; this eliminates the possibility of sharp folds. The first von K{\'a}rm{\'a}n equation (in a distributional form) follows from \eqref{N1IncompatibilityElasticStrain1}, after writing $\boldsymbol{E}^e$ in terms of $\boldsymbol{\Sigma}$ and then $\Phi$, and substituting ${\boldsymbol{\Lambda}}$ in terms of $W$, as 
\begin{equation}
\frac{1}{E}\Delta^2 \Phi +\frac{1}{2} [W,W]=- N_1 + \Det(\boldsymbol{\Lambda}^p). \label{vk1pdist1}
\end{equation}
The equivalent local form of  \eqref{vk1pdist1} can be obtained using Identity~\ref{LaplacianFormula} as
\begin{subequations}
\label{vk1ploc}
\begin{align}
\frac{1}{E}\Delta^2 \phi +\frac{1}{2} [w,w]=-\eta + \det(\boldsymbol{\lambda}^p) ~ \text{in}~\Omega-\{S\cup\{O\}\}, \label{vk1ploc1}
\\
\frac{1}{E}\left\langle \llbracket \div \nabla\nabla \phi \rrbracket, \boldsymbol{\nu} \right\rangle + \frac{1}{E}\left\langle \frac{d}{ds} \llbracket\nabla\nabla \phi\rrbracket,\boldsymbol{t}\otimes \boldsymbol{\nu} \right\rangle = 
 -\zeta_1 ~ \text{on}~S-\{O\}, 
 \label{vk1ploc2}
\\
\frac{1}{E}\left\langle \llbracket\nabla\nabla \phi\rrbracket, \boldsymbol{\nu}\otimes \boldsymbol{\nu} \right\rangle=-\zeta_2 ~\text{on} ~S-\{O\},~\text{and} \label{vk1ploc3}
\\
\frac{1}{E}\Phi(\Delta^2 v^\alpha)-\nabla W\otimes \nabla W(\mathbb{A}\nabla\nabla v^\alpha)=-N_1(v^\alpha) ~\text{at}~O, \label{vk1ploc4}
\end{align}
\end{subequations}
for all multi indices $\alpha \in \mathbb{N}^2$ such that $|\alpha| \leq q$, where $q$ is the maximum of $\deg(\Phi)+4$, $\deg(\nabla W\otimes \nabla W)+2$, and $\deg(N_1)$. The equivalency of \eqref{vk1ploc} and \eqref{vk1pdist1}  can be established using Lemma~\ref{RepresentationELemma}. The second von K{\'a}rm{\'a}n equation (in a distributional form) is as given in \eqref{vk2dist1} whose equivalence to the following local form can be established using \cite[Id. 2.2]{pandey2020topological}, Identity~\ref{CurlCurlFormulas}, Identity~\ref{LaplacianFormula}, and Lemma~\ref{RepresentationELemma}:
\begin{subequations}
\label{VonKArmanEquilibriumStrongForms1}
\begin{align}
D\Delta^2 w - [\phi,w]=f_1 + D\left( (1-\nu)\div \div {\boldsymbol{\lambda}^p} +\nu \Delta (\tr({\boldsymbol{\lambda}^p}))\right) ~ \text{in}~\Omega-\{S\cup\{O\}\},
\\
D \left\langle \llbracket \div \nabla\nabla w \rrbracket, \boldsymbol{\nu} \right\rangle  = f_2 + D(1-\nu) \langle \llbracket \div \boldsymbol{\lambda}^p \rrbracket, \boldsymbol{\nu} \rangle  + D\nu\llbracket \nabla(\tr \boldsymbol{\lambda}^p) \rrbracket   ~\text{on} ~S-\{O\},
\\
 \llbracket\Delta w\rrbracket = 0  ~\text{on} ~S-\{O\},~\text{and}
\\
D W(\Delta^2 v^\alpha)+\nabla \Phi \otimes \nabla W(\mathbb{A}\nabla\nabla v^\alpha)=F(v^\alpha)-D\left( (1-\nu)\boldsymbol{\Lambda}^p(\nabla\nabla v^\alpha) +\nu  (\tr(\boldsymbol{\Lambda}^p)(\Delta v^\alpha))\right)~\text{at}~O,
\end{align}
\end{subequations}
for all multi indices $\alpha \in \mathbb{N}^2$ such that $|\alpha| \leq q$, where $q$ is the maximum of $\deg(W)+4,$ $\deg(\nabla \Phi\otimes \nabla W)+2$, $\deg(F)$, and $\deg(\boldsymbol{\Lambda}^p)+2$. The complete set of the desired von K{\'a}rm{\'a}n equations (in Case 2) is given as distributional equations in \eqref{vk1pdist1} and \eqref{vk2dist1} or, equivalently, as local pointwise equations in \eqref{vk1ploc} and \eqref{VonKArmanEquilibriumStrongForms1}. We end by noting that there is an alternate scenario where the generalized determinant in \eqref{vk1pdist1} is well defined but ${\boldsymbol{\Lambda}^p}$ is not necessarily continuous. This is possible if we assume $\Curl {\boldsymbol{\Lambda}^p} = \boldsymbol{0}$, i.e., there exist a distribution $W^p \in \mathcal{D}(\Omega)$ such that $\boldsymbol{\Lambda}^p=\nabla\nabla W^p$, and $W^p$ satisfies the assumptions mentioned in \ref{a2p}.

\section{Applications} \label{applications}

\subsection{Conical deformation} \label{conical}
A transverse deformation of the form $w=rg(\theta)$ is referred to as a conical deformation (with the tip of the cone at a point $O \in \Omega$). Such deformations appear as solutions in the problems of isolated disclinations~\cite{SeungNelson88, PandeyPRE} and crumpled sheets~\cite{ben1997crumpled, cerda1998conical, witten2007}. We establish two results (as two lemmas) in the context of conical deformations and discuss their implications for inextensible (i.e., $\boldsymbol{E}^e=\boldsymbol{0}$) F{\"o}ppl-von K{\'a}rm{\'a}n plate domains with fields having a singular support as a subset of $\{O\}$ (hence, $S = \emptyset$). The first is a kinematical result where we seek a general form of the transverse deformation for a compatible bending strain which has a singular support at $O$ and a Gaussian curvature which is necessarily trivial outside $O$. As an immediate consequence of the result we show that the Gaussian curvature at $O$ is a Dirac, and hence dislocation and vacancy like defects are not supported. The second result considers a specific form of the deformation and the stress function and shows that no point force (or couple) can appear at $O$ as a result of transverse force equilibrium. The implications of these results on surfaces with an isolated defect and surfaces under constrained deformation will be noted. 
\begin{lemma}
\label{GaussianCurvatureDiracLemma}
Let $\boldsymbol{\Lambda} \in \mathcal{D}'(\mathbb{R}^2, \Sym)$ be such that $\deg(\boldsymbol{\Lambda})<0$, $\singsupp(\boldsymbol{\Lambda})= \{O\}$, and $\boldsymbol{\Lambda}|_{\mathbb{R}^2-O} (\boldsymbol{x}) \neq \boldsymbol{0}$ for all $\boldsymbol{x}\in \mathbb{R}^2-O$. The following statements are equivalent:
\begin{enumerate}[align = left]
\item[\textup{i.}] $\Det \boldsymbol{\Lambda}|_{\mathbb{R}^2-O}=0$ and $\Curl \boldsymbol{\Lambda}=\boldsymbol{0}$.
\item[\textup{ii.}] $\boldsymbol{\Lambda}(\boldsymbol{\psi})=\int_{\mathbb{R}^2}\left\langle({f(\theta)}/{r}) \boldsymbol{e_\theta}\otimes \boldsymbol{e_\theta}, \boldsymbol{\psi} \right\rangle \da$ for all $\boldsymbol{\psi}\in \mathcal{D}(\mathbb{R}^2,\Lin)$ such that $\int_0^{2\pi}f(\theta)\boldsymbol{e_\theta}d\theta=\boldsymbol{0}$. 
\item[\textup{iii.}]  There exists $W \in \mathcal{D}'(\Omega)$ such that $W(\psi)=\int_{\mathbb{R}^2}w\psi \da$ for all $\psi \in \mathcal{D}(\mathbb{R}^2)$, where $w=rg(\theta)$ and $g+{d^2g}/{d\theta^2}=f$, satisfying $\boldsymbol{\Lambda}=\nabla\nabla W$.
\end{enumerate}
\begin{proof} (ii.$\implies$i.) A direct computation of $\Curl \boldsymbol{\Lambda}$ and $\Det \boldsymbol{\Lambda}|_{\mathbb{R}^2-O}$ for the given form of $\boldsymbol{\Lambda}$ in ii. yields the desired result. 
(i.$\implies$ii.) Given $\Det \boldsymbol{\Lambda}|_{\mathbb{R}^2 - O}=0,$ there exists a unit vector field $\boldsymbol{v}_1 \in C^\infty (\mathbb{R}^2-O,\mathbb{R}^2)$, expressible as $\boldsymbol{v}_1=-\sin \alpha \boldsymbol{e}_1+\cos \alpha \boldsymbol{e}_2$ where $\alpha \in C^\infty(\mathbb{R}^2-O,[0,2\pi))$, and $F\in C^\infty(\mathbb{R}^2-O)$ such that $\boldsymbol{\lambda}=F \boldsymbol{v}_1\otimes \boldsymbol{v}_1$ with $\boldsymbol{\Lambda}(\boldsymbol{\psi})=\int_{\mathbb{R}^2}\left\langle\boldsymbol{\lambda}, \boldsymbol{\psi} \right\rangle \da$ for all $\boldsymbol{\psi}\in \mathcal{D}(\mathbb{R}^2,\Lin)$. Let $\boldsymbol{v_2}=\cos \alpha \boldsymbol{e}_1+\sin \alpha \boldsymbol{e}_2$. In $\mathbb{R}^2-O$, $\curl \boldsymbol{\lambda}=(-\langle \nabla F,\boldsymbol{v}_2 \rangle+F \langle\nabla \alpha, \boldsymbol{v}_1 \rangle) \boldsymbol{v}_1 + F\langle \nabla \alpha, \boldsymbol{v}_2 \rangle \boldsymbol{v}_2$. Hence $\curl \boldsymbol{\lambda}=\boldsymbol{0}$ implies $\langle \nabla \alpha, \boldsymbol{v}_2\rangle=0$ and $\langle \nabla F, \boldsymbol{v}_2\rangle=F \langle\nabla \alpha, \boldsymbol{v}_1 \rangle$. According to the former, the integral curves of the vector field $\boldsymbol{v}_2$ are straight lines each with a fixed value of $\alpha$. The point of singularity in $\boldsymbol{\Lambda}$, i.e., $O$, is necessarily at the intersection of two integral curves $\boldsymbol{v}_2$. Consequently, the integral curves of $\boldsymbol{v}_2$ are radial lines intersecting at the point O such that $\alpha=\theta$ in $\mathbb{R}^2-O$. Moreover, we then have $\boldsymbol{\lambda}=F \boldsymbol{e}_\theta \otimes \boldsymbol{e}_\theta$ such that ${\partial F}/{\partial r}={F}/{r}$ which yields $F={f(\theta)}/{r}$. That $\int_0^{2\pi}f(\theta)\boldsymbol{e}_\theta d\theta=\boldsymbol{0}$ follows immediately from $\Curl \boldsymbol{\Lambda}=\boldsymbol{0}$.
(ii.$\implies$iii.) Given $f$, we introduce $g$ such that $g+{d^2g}/{d\theta^2}=f$. The existence of a $W$, such that $W(\psi)=\int_{\mathbb{R}^2}w\psi \da$ for all $\psi \in \mathcal{D}(\mathbb{R}^2)$, satisfying $\boldsymbol{\Lambda} = \nabla\nabla W$ follows from Lemma~\ref{SymFieldsAiryHessian}. That $w=rg(\theta)$ can be verified by a direct substitution.
(iii.$\implies$ii.) A direct computation of $\boldsymbol{\Lambda} = \nabla\nabla W$ for the given form of $W$ yields the required result.
\end{proof}
\end{lemma}

It is important to note that the above lemma assumes an unbounded domain, i.e., we take $\Omega$ as $\mathbb{R}^2$. This is necessary since otherwise the integral curves of $\boldsymbol{v}_2$ can possibly intersect outside $\Omega$. There can be many such intersections outside $\Omega$ and only one within $\Omega$ at $O$ such that the singular support of $\boldsymbol{\Lambda}$ remains at $O$. This however would yield a rich class of solutions beside $w=r g(\theta)$, something we do not pause to discuss presently.

To evaluate  the Gaussian curvature $\Det(\boldsymbol{\Lambda})$, corresponding to the form of displacement $W$ in Lemma~\ref{GaussianCurvatureDiracLemma}, we begin by calculating $\nabla w= g\boldsymbol{e}_r+({dg}/{d\theta})\boldsymbol{e}_\theta$ and consequently $\nabla w \otimes \nabla w=g^2\boldsymbol{e}_r\otimes \boldsymbol{e}_r+g({dg}/{d\theta})(\boldsymbol{e}_r\otimes \boldsymbol{e}_\theta+\boldsymbol{e}_\theta\otimes \boldsymbol{e}_r)+({dg}/{d\theta})^2 \boldsymbol{e}_\theta\otimes\boldsymbol{e}_\theta$. As a result, we have
\begin{equation}
\Curl (\nabla W \otimes \nabla W) (\boldsymbol{\psi})=\int_\Omega -\frac{g}{r} \left(g+\frac{d^2g}{d\theta^2}\right)\langle\boldsymbol{e}_\theta,\boldsymbol{\psi} \rangle\da 
\end{equation}
for all $\boldsymbol{\psi} \in \mathcal{D}(\Omega,\mathbb{R}^2)$. For any $\boldsymbol{V}\in \mathcal{D}'(\Omega,\mathbb{R}^2)$, such that $\boldsymbol{V}(\boldsymbol{\psi})=\int_\Omega ({h(\theta)}/{r})\langle\boldsymbol{e}_\theta,\boldsymbol{\psi} \rangle\da$, 
$\Curl \boldsymbol{V}=\left( \int_0^{2\pi} h(\theta) d\theta \right) \delta_O$. Taking $\boldsymbol{V}$ to be $\Curl (\nabla W \otimes \nabla W)$, we obtain
\begin{equation}
\Curl \Curl (\nabla W \otimes \nabla W)= - \left( \int_0^{2\pi} g\left(g+\frac{d^2g}{d\theta^2}\right) d\theta \right) \delta_O.
\end{equation}
Recalling that $\Det(\boldsymbol{\Lambda})=-({1}/{2}) \Curl \Curl (\nabla W \otimes \nabla W)$ gives us the required formula:
\begin{equation}
\Det (\boldsymbol{\Lambda})=\frac{1}{2}\left(\int_0^{2\pi}g \left(g+\frac{d^2g}{d\theta^2} \right)d\theta\right) \delta_O. \label{detlamb1}
\end{equation}
Furthermore,  Since $g(\theta)$ is necessarily periodic with a period of $2\pi$ we have the representation $g(\theta)=\sum_{k=0}^{\infty} g_k \sin(k\theta)$. Substituting this into \eqref{detlamb1} we obtain
\begin{equation}
\Det (\boldsymbol{\Lambda})=\left(\pi {g_0}^2-\frac{1}{2}\sum_{k=0}^{\infty} {g_k}^2(k^2-1)\right) \delta_O. \label{detlamb2}
\end{equation}
The point supported Gaussian curvature $\Det (\boldsymbol{\Lambda})$ therefore is necessarily a Dirac for conical distributions. All the other terms in the general representation of a
point supported $\Det (\boldsymbol{\Lambda})$, such as those present in \eqref{representationE}, are absent. The strength of the Dirac singularity at the tip of the cone is $({1}/{2})  \int_0^{2\pi} g(g+{d^2g}/{d\theta^2}) d\theta$. For a developable cone~\cite{cerda1998conical}, which is characterized through a conical deformation with trivial generalized Gaussian curvature, $({1}/{2}) \int_0^{2\pi} g(g+{d^2g}/{d\theta^2}) d\theta=0$. On the other hand, for the inextensible plate domain with an isolated defect at $O$, we have
$\Det (\boldsymbol{\Lambda})=-({\Theta} + \Curl \boldsymbol{A} + \Curl \Curl \boldsymbol{Q})$ (recall \eqref{N1IncompatibilityElasticStrain1} and \eqref{IncompatibilityEquation1}), where the disclination density ${\Theta}$, the dislocation density $\boldsymbol{A}$, and the density of point defects $\boldsymbol{Q}$ are Dirac distributions. Clearly, an isolated disclination is permissible with strength $s= ({1}/{2}) \int_0^{2\pi} g(g+{d^2g}/{d\theta^2}) d\theta$. However, neither dislocations nor point defects are allowed. Indeed, if we consider an an isolated dislocation (of Burgers vector $\boldsymbol{b}$)  at $O$, i.e., $\boldsymbol{A}= \boldsymbol{b}\delta_O$, then the resulting incompatibility is $N_1=\langle (\boldsymbol{b}\times \boldsymbol{e}_3),\nabla \delta_O\rangle$, which in turn would induce a dipole like source term (at $O$) for the Gaussian curvature. This indicates that one would need to incorporate elastic extensibility of the plate, or look beyond conical deformations, in order to consider isolated defects besides disclinations.

Under the inextensibility constraint, the stress function $\Phi$ is no longer determined constitutively; it is interpreted as the Lagrange multiplier associated with the constraint. The first term (on the left hand side) of the (generalized) first von K{\'a}rm{\'a}n equation (\eqref{vk1dist1} or \eqref{vk1pdist1}) vanishes and the second term is of the form \eqref{detlamb1}, or \eqref{detlamb2}, under the present considerations. The stress function appears only in the (generalized) second von K{\'a}rm{\'a}n equation \eqref{vk2dist1}. 
In the following lemma, we show that a form of conical deformation and stress function satisfies the equilibrium condition given through  \eqref{vk2dist1} with a trivial externally applied body force field.

\begin{lemma}
\label{gthetalambdaequilibriumSatisfactionLemma}
Given $W\in \mathcal{D}'(\Omega)$ such that $W(\psi)=\int_\Omega w\psi \da$ for all $\psi \in \mathcal{D}(\Omega)$, where $w=rg(\theta)$ with $g(\theta)=a_{\mu}\sin(\mu \theta)$, and $\Phi\in \mathcal{D}'(\Omega)$ such that $\Phi(\psi)=\int_\Omega \phi \psi \da$ for all $\psi \in \mathcal{D}(\Omega)$, where $\phi=D\lambda \ln r$ with $\lambda=\mu^2 - 1$, they satisfy $D\Delta^2 W - [\Phi,W]=0$. 
\begin{proof}
The equation $D\Delta^2 W - [\Phi,W]=0$ can be equivalently written as 
\begin{equation}
D\Div\Div (\nabla\nabla W)+\Div\Div((\boldsymbol{e}_3 \times\nabla {\Phi})\otimes(\boldsymbol{e}_3 \times\nabla {W}))=0. \label{vk2dist10}
\end{equation}
For the given forms of $W$ and $\Phi$ the restriction of  \eqref{vk2dist10} to $\Omega-O$ yields 
\begin{equation}
\label{gthetalambdaequilibrium}
\frac{d^4g}{d\theta^4}+(\lambda+2)\frac{d^2g}{d\theta^2}+(\lambda+1)g=0,
\end{equation}
which is identically satisfied by $g(\theta)=a_{\mu}\sin(\mu \theta)$ with $\lambda=\mu^2-1$. To ensure that $D\Delta^2 W - [\Phi,W]=0$ is satisfied we have to additionally check that \eqref{LoopIntegralPoinForceDipole} holds (with vanishing right hand side). Towards this end, let $\boldsymbol{A}=(\nabla\nabla W)+((\boldsymbol{e}_3 \times\nabla {\Phi})\otimes(\boldsymbol{e}_3 \times\nabla {W})).$ Then, for the given forms of $W$ and $\Phi$, $\boldsymbol{A}(\boldsymbol{\psi})=\int_\Omega \langle\boldsymbol{a},\boldsymbol{\psi}\rangle\da$ for all $\boldsymbol{\psi}\in \mathcal{D}(\Omega,\Lin)$, where $\boldsymbol{a}=(1/r) \left( (g+{d^2g}/{d\theta^2} + \lambda g)\boldsymbol{e}_\theta\otimes \boldsymbol{e}_\theta -{\lambda {dg}/{d\theta}} \boldsymbol{e}_\theta \otimes \boldsymbol{e}_r \right)$. Further calculations yield $\div \boldsymbol{a}=(1/r^2) \left( ({d^3g}/{d\theta^3}+(\lambda+1){dg}/{d\theta}) \boldsymbol{e}_\theta- ({d^2g}/{d\theta^2}+(\lambda+1)g)\boldsymbol{e}_r \right)$. The desired result follows on substituting the  expressions for $\boldsymbol{a}$ and $\div \boldsymbol{a}$ in \eqref{LoopIntegralPoinForceDipole} with $g(\theta)=a_{\mu}\sin(\mu \theta)$ and $\lambda=\mu^2-1$.
\end{proof}
\end{lemma}

\begin{figure}[t!]
\begin{subfigure}{.49\textwidth}
  \centering
\includegraphics[scale=0.3]{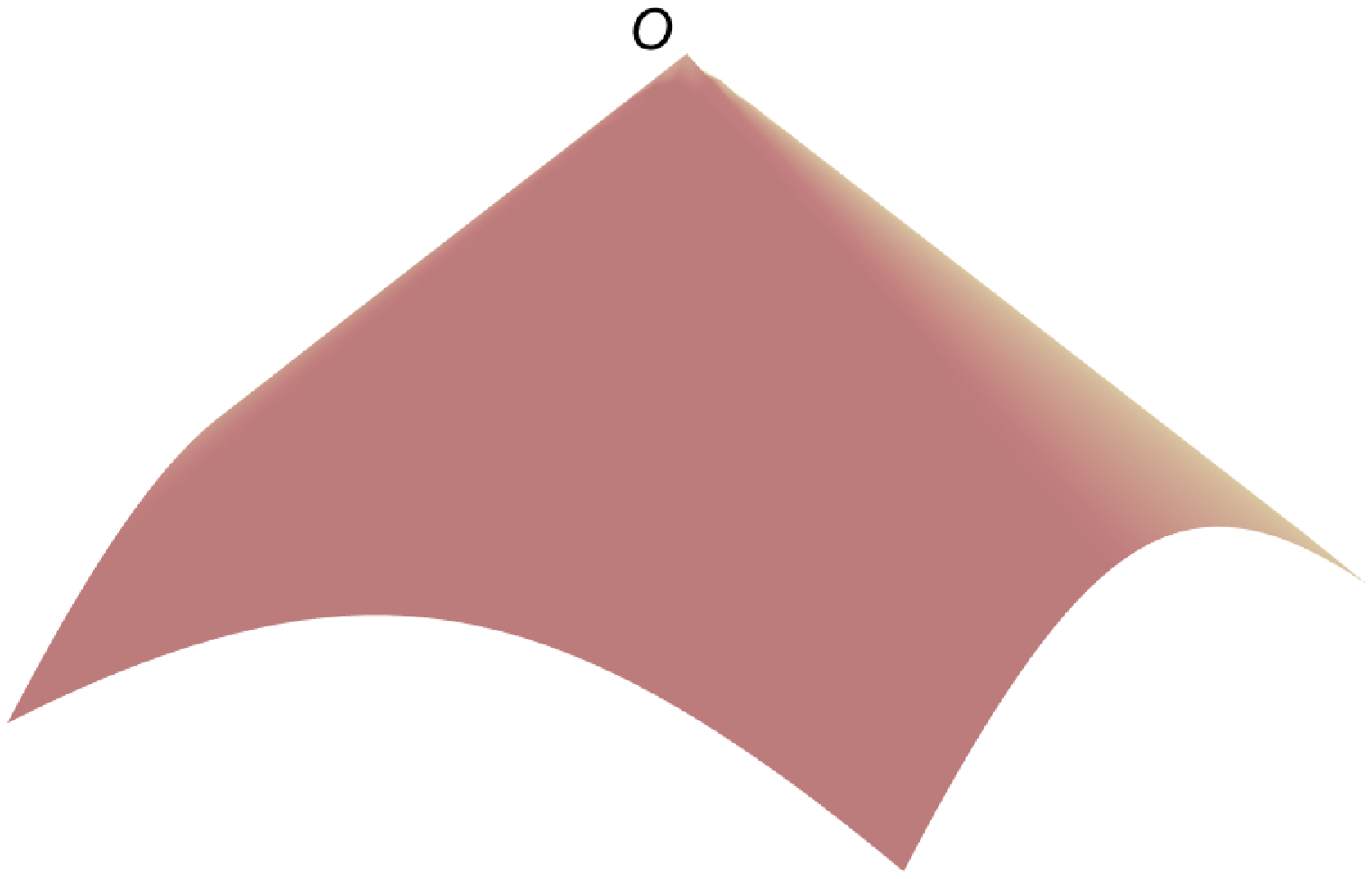}
\subcaption{Positive disclination.} \label{posdisfig}
\end{subfigure}%
%\hspace{35pt}
\begin{subfigure}{.49\textwidth}
  \centering
\includegraphics[scale=0.3]{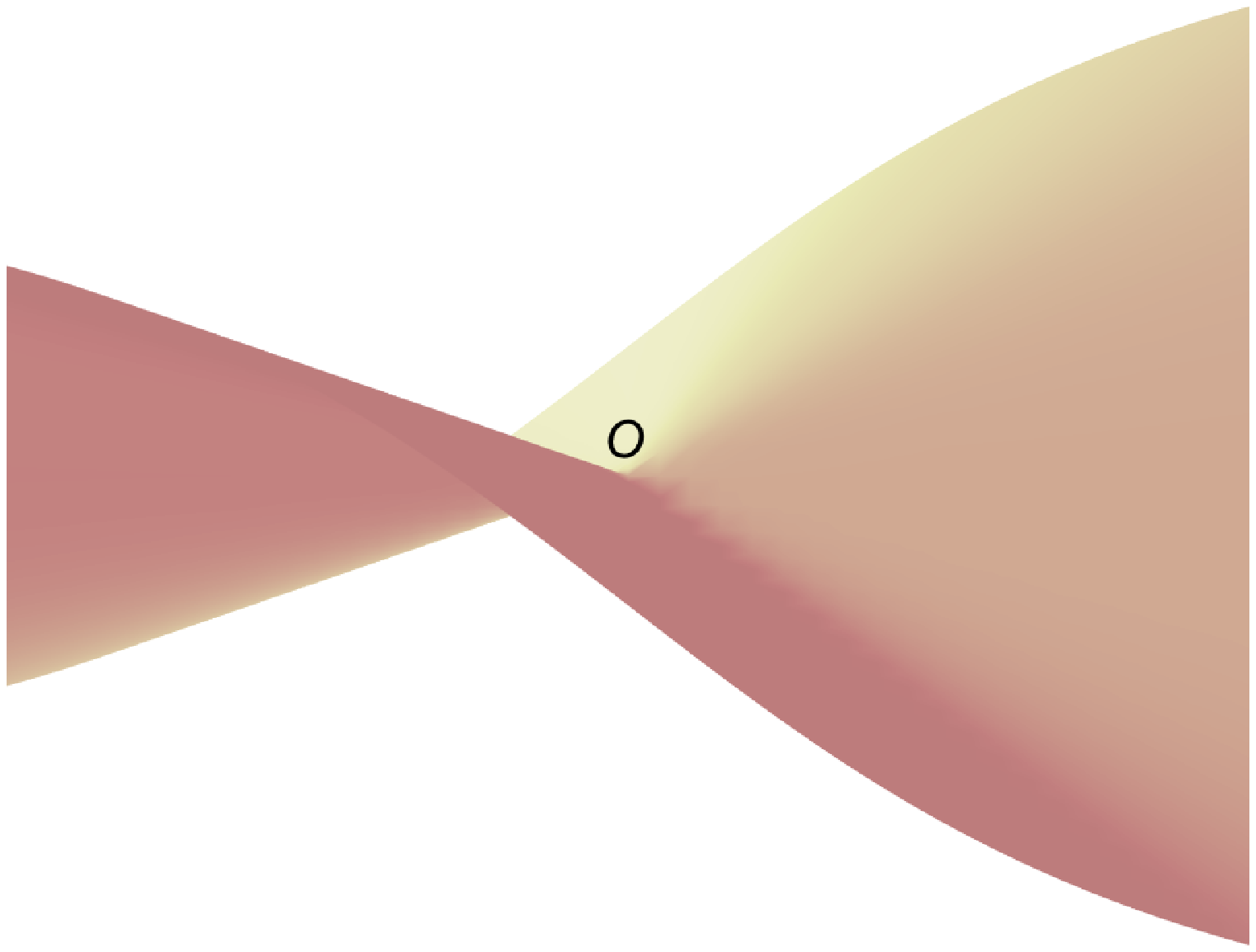}
\subcaption{Negative disclination.} \label{negdisfig}
\end{subfigure}%
\caption{The transverse deformation of a flat, elastically inextensible, sheet in the presence of an isolated disclination at $O$.}
\label{isodisclfigs}
\end{figure}

The preceding lemma, in conjunction with \eqref{detlamb2}, can be used to obtain a solution to the disclination problem.
 We consider an isolated point disclination, i.e., $\Theta=-s\delta_O$, where $s$ is the strength of the disclination (or the disclination charge). Then, in the absence of plastic bending strains and elastic inextensibility, we have $\Det(\boldsymbol{\Lambda})=s\delta_O$. A comparison with \eqref{detlamb2} yields
\begin{equation}
s=\pi {g_0}^2 - \frac{\pi}{2} \sum {g_k}^2(k^2-1). \label{disclcharge}
\end{equation}
Equation \eqref{disclcharge} is a geometrical relation which captures the balance of the disclination charge with the Gaussian curvature of the conical deformation. If there is no disclination (i.e., we have a developable cone) then $s=0$ and  \eqref{disclcharge} ensures that the conical tip has a vanishing Gaussian curvature. A solution of the positive disclination problem ($s>0$) is obtained if we identify $g_0=\sqrt{{s}/{\pi}}$ (i.e., $\mu=0$) and take  $\phi=-D\ln r$ (i.e., $\lambda = -1$). Similarly, a solution of the negative disclination problem ($s<0$) is obtained if we identify $g_2=\sqrt{{-2s}/{3\pi}}$ (i.e., $\mu=2$) and take  $\phi=3D\ln r$ (i.e., $\lambda = 3$). According to Lemma~\ref{gthetalambdaequilibriumSatisfactionLemma}, these solutions satisfy the equilibrium equations with no external forces. The deformations are illustrated in Figure~\ref{isodisclfigs}.

Recall (from \eqref{MomentBalance1}) that the equilibrium condition, away from the tip of the cone with trivial transverse forces, is  $D\Delta^2 w-\langle\boldsymbol{\sigma},\boldsymbol{\lambda}\rangle=0$. For a conical deformation this takes the form 
\begin{equation}
\frac{1}{r^3} \left( g+2\frac{d^2g}{d\theta^2}+\frac{d^4g}{d\theta^4} \right)-\frac{\sigma_{\theta \theta}}{r}\left(g+\frac{d^2g}{d\theta^2}\right)=0,
\end{equation}
where $\sigma_{\theta \theta}=\langle \boldsymbol{\sigma},\boldsymbol{e}_\theta\otimes \boldsymbol{e}_\theta \rangle$. This is equivalent to \eqref{gthetalambdaequilibrium} once we identify $\sigma_{\theta \theta}=-{\lambda}/{r^2}$.  An alternate way to obtain \eqref{gthetalambdaequilibrium} is to  minimize the total bending energy (away from the cone tip)
\begin{equation}
U_b=\int_{\Omega-B_\epsilon} k_m \frac{1}{r^2} \left( g + \frac{d^2g}{d\theta^2}\right)^2 \da
\end{equation}
 with respect to $g$~\cite{cerda1998conical}, where $k_m$ is a uniform material constant, under the constraint on the Gaussian curvature at the tip of the cone, $ \int_0^{2\pi} g(g+{d^2g}/{d\theta^2}) d\theta =2s$; $\lambda$ appears as the Lagrange multiplier corresponding to the constraint. The energetic framework implicitly assumes $\sigma_{\theta\theta}=-{\lambda}/{r^2}$. It is not straightforward to incorporate a more general stress field in the existing energetic framework. Moreover note that, since the energy based framework is predicated on ignoring the vicinity of the singularity, we do not obtain any equilibrium condition at the point of singularity and hence forces can not be prescribed at the tip of the cone. In our framework, a point force and a point force dipole can be prescribed at the singular point through the integral balance law \eqref{LoopIntegralPoinForceDipole}.

Finally, we briefly discuss the problem of constrained deformation where the sheet is unilaterally constrained to adhere to an externally imposed rigid shape. An example is afforded by the problem of squeezing a sheet of paper inside a hollow sphere (or inside our fist). The sheet responds by crumpling and developing a distribution of kinks and creases~\cite{witten2007}. A simpler example is to try to fit a sheet of paper inside a hollow cone~\cite{cerda1998conical}. This yields a developable cone (conical singularity with $s=0$) such that there are portions of the of the sheet directly in contact with the outer rigid cone and portions which are folded away. A conical deformation in such a scenario would satisfy the equilibrium of the form $D\Delta^2 W - [\Phi,W]=B$ where $B$ is the distributional contact force supported over the region where the sheet is in contact with the rigid cone. If the constrained deformation has been achieved by applying a point force then the singular force should be incorporated within $B$.

\subsection{Folds} \label{folds}

In this section, we assume the domain $\Omega$ to be without any point $O$ of singularity and having a singular interface $S$ which is a smooth curve such that $\partial S - \partial \Omega = \emptyset$. We assume both the bulk and the interfacial fields, and their derivatives, to remain bounded in the entire domain. We restrict our attention to elastically inextensible plates (i.e., $\boldsymbol{E}^e = \boldsymbol{0}$ in $\Omega$). We define a fold as an interfacial concentration in plastic bending strain. For the present discussion we will assume that there is no bulk contribution to the plastic bending strain, i.e., $\boldsymbol{\Lambda}^p \in \mathcal{C}(\Omega,\Sym)$ with a smooth line density $\boldsymbol{\gamma}$ such that $\boldsymbol{\Lambda}^p (\boldsymbol{\psi})=\int_S \langle \boldsymbol{\gamma},\boldsymbol{\psi} \rangle \dl$ for all $\boldsymbol{\psi} \in \mathcal{D}(\Omega,\Lin)$. The compatibility condition $\boldsymbol{\gamma}\times \boldsymbol{\nu}=\boldsymbol{0}$ implies that $\boldsymbol{\gamma}$ is of the form $\boldsymbol{\gamma}=\gamma_0 \boldsymbol{\nu}\otimes\boldsymbol{\nu}$. The fold is therefore prescribed in terms of a smooth scalar field $\gamma_0$ (the strength of the fold) and the geometry of the curve $S$. Moreover
\begin{equation}
\llbracket \nabla w \rrbracket=-\gamma_0 \boldsymbol{\nu}\otimes \boldsymbol{\nu}. \label{folddispjump}
\end{equation}
With the above considerations, and under the assumption of elastic inextensibility, the first von K{\'a}rm{\'a}n equation \eqref{vk1loc} reduces to
\begin{subequations}
\label{foldcomp}
\begin{align}
\label{FoldCylindricalBulkCompatibility}
\frac{1}{2}[w,w]=-\curl\curl \boldsymbol{e}^p~\text{in}~\Omega-S,
\\
\label{FoldCylindricalInterfacialCompatibility1}
\gamma_0 \langle \{ \nabla\nabla w \},\boldsymbol{t}\otimes\boldsymbol{t} \rangle = \langle \llbracket \nabla \boldsymbol{e}^p \rrbracket, \boldsymbol{r} \rangle + k\langle \llbracket \boldsymbol{e}^p \rrbracket, \boldsymbol{\nu} \otimes \boldsymbol{\nu} \rangle ~\text{on}~S,~\text{and}
\\
\label{FoldCylindricalInterfacialCompatibility2}
\langle \llbracket \boldsymbol{e}^p \rrbracket, \boldsymbol{t} \otimes \boldsymbol{t} \rangle=0 ~\text{on}~S,
\end{align}
\end{subequations}
while  the second von K{\'a}rm{\'a}n equation \eqref{VonKArmanEquilibriumStrongForms}, in the absence of external forces, takes the form
\begin{subequations}
\label{EquilibriumEq}
\begin{align}
\label{FoldBulkEquilibrium}
D\Delta^2w - [\phi,w]=0~\text{in}~\Omega-S,
\\
\label{FoldInterfacialEquilibrium1}
-D(1-\nu)\frac{d^2 \gamma_0}{ds^2}+D\langle \llbracket \div \nabla\nabla w \rrbracket,\boldsymbol{\nu} \rangle + \gamma_0 \langle  \nabla\nabla \phi , \boldsymbol{t} \otimes \boldsymbol{t} \rangle=0 ~\text{on}~S,~\text{and}
\\
\label{FoldInterfacialEquilibrium2}
\llbracket \Delta w \rrbracket = (1-\nu)k \gamma_0  ~\text{on}~S.
\end{align}
\end{subequations}

\begin{figure}[t!]
\begin{subfigure}{.49\textwidth}
  \centering
\includegraphics[scale=0.3]{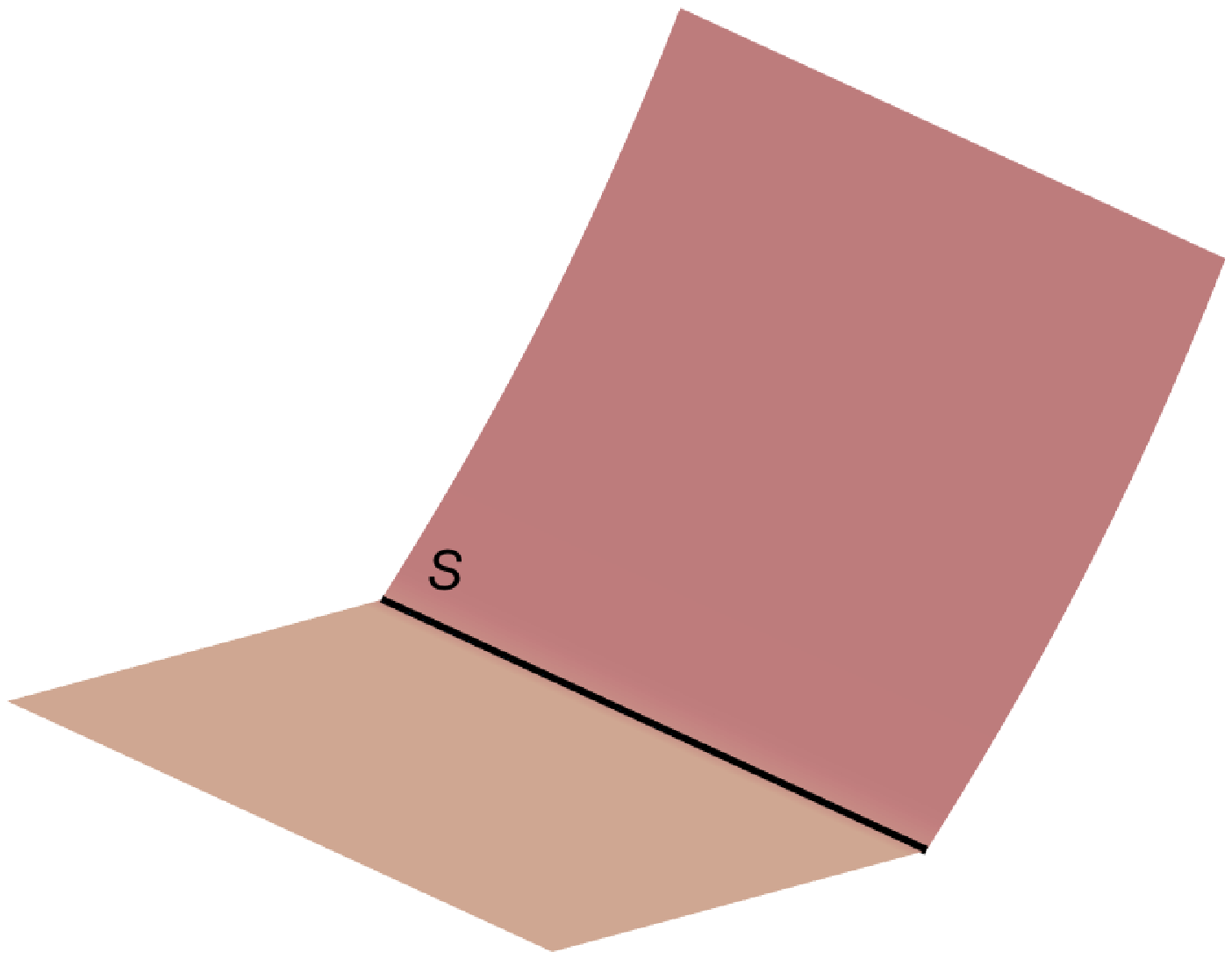}
\subcaption{A linear fold.} \label{linfold}
\end{subfigure}%
%\hspace{35pt}
\begin{subfigure}{.49\textwidth}
  \centering
\includegraphics[scale=0.3]{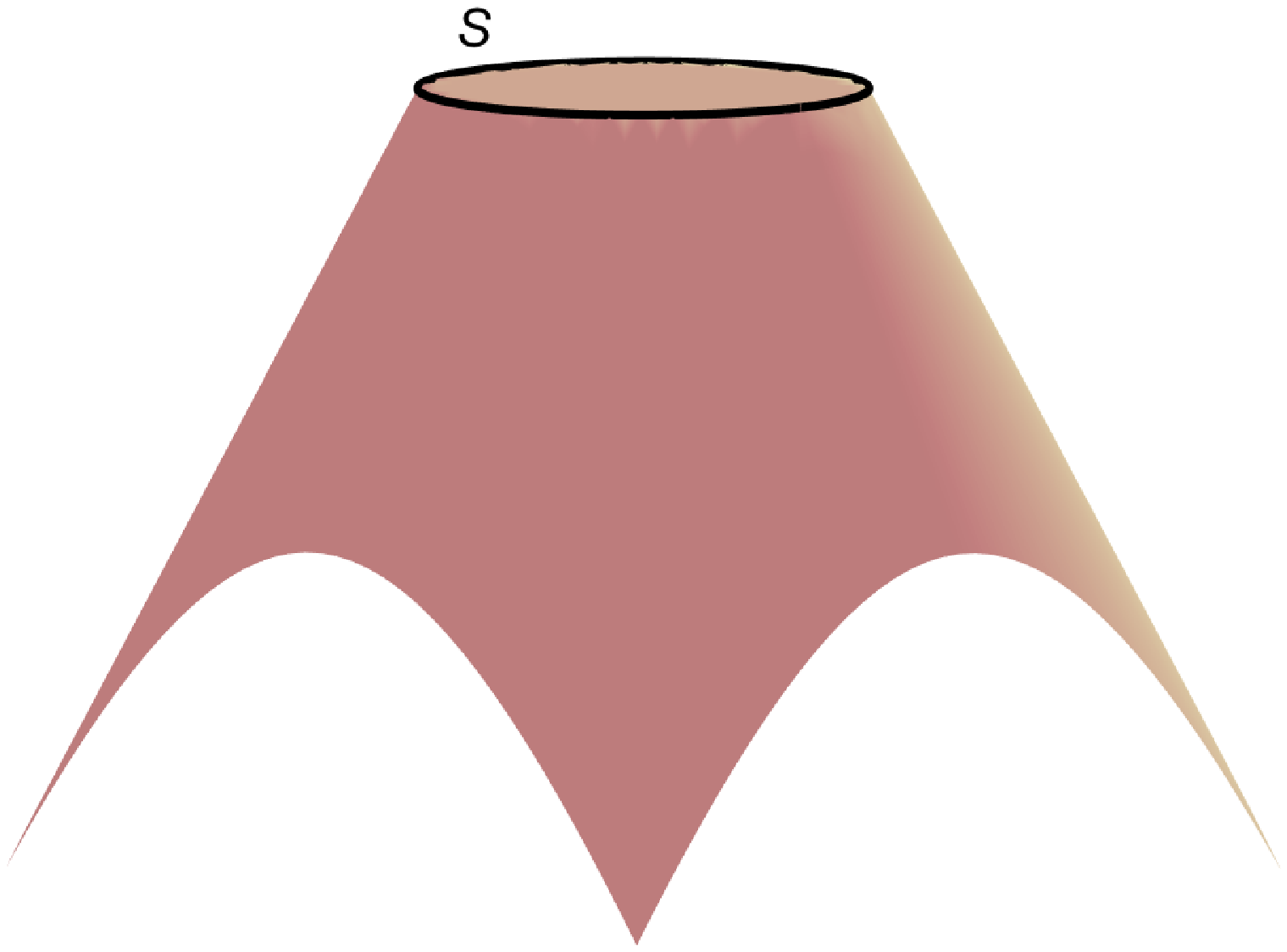}
\subcaption{A circular fold.} \label{cirfold}
\end{subfigure}%
\caption{The transverse deformation of a flat, elastically inextensible, sheet in the presence of a linear and a circular fold. In the former, curvature on one side of the fold develops in response to the moment and transverse force applied at the edge. In the latter, a distribution of transverse force couples at the fold location is required to maintain the desired shape of the deformation. The fold $S$ is indicated with a bold black line.}
\label{twofolds}
\end{figure}

We discuss the nature of these two sets of equations in the context of two problems of folding an elastic sheet along a singular curve. First, we consider a fold along a straight line $S$ (hence $\boldsymbol{\nu}$ is fixed along $S$) such that the deformation on either side of the fold is cylindrical, i.e., $\boldsymbol{\lambda}=g(\langle\boldsymbol{x},\boldsymbol{c}\rangle) \boldsymbol{c}\otimes\boldsymbol{c}$, where $\boldsymbol{c}$ is a constant vector; see Figure~\ref{linfold}. A transverse deformation which satisfies $\boldsymbol{\lambda} = \nabla \nabla w$ (outside $S$) is of the form $w=f(\langle\boldsymbol{x},\boldsymbol{c}\rangle)$ such that ${d^2 f}/{d{q}^2}=g$, where $q=\langle \boldsymbol{x}, \boldsymbol{c} \rangle$. Let $q=0$ represent $S$. At the fold $S$, \eqref{folddispjump} implies $\llbracket {df}/{d{q}} \rrbracket\boldsymbol{c}=-\gamma_0 \boldsymbol{\nu}$, from which we conclude that the interfacial line is necessarily orthogonal to $\boldsymbol{c}$, i.e., $\langle \boldsymbol{t},\boldsymbol{c}\rangle=0$ and $\llbracket {df}/{d{q}} \rrbracket=-\gamma_0$. Accordingly, ${d \gamma_0}/{d {s}}=0$, i.e., $\gamma_0$ is necessarily constant. We also assume a vanishing plastic stretching strain (i.e., $\boldsymbol{e}^p=\boldsymbol{0}$). The assumed cylindrical form of the deformation then satisfies the bulk compatibility conditions~\eqref{foldcomp}. We can use the coordinates $(q,s)$ to describe $\Omega$ (even outside $S$). The equilibrium equations \eqref{EquilibriumEq} reduce to
\begin{subequations}
\begin{align}
\label{FoldCylindricalBulkEquilibrium}
D\frac{d^4 f}{d{q}^4}-\frac{\partial^2 \phi}{\partial{s}^2}\frac{d^2 f}{d{q}^2}=0~\text{in}~\Omega-S,
\\
\label{FoldCylindricalInterfacialEquilibrium1}
D \left \llbracket \frac{d^3 f}{d{q}^3}  \right \rrbracket + \gamma_0 \frac{\partial^2\phi}{\partial s^2}=0 ~\text{on}~S,~\text{and}
\\
\label{FoldCylindricalInterfacialEquilibrium2}
\left\llbracket \frac{d^2 f}{d{q}^2} \right\rrbracket = 0  ~\text{on}~S.
\end{align}
\end{subequations}
 The deformation $f(q)= \gamma_0 q$, if $q \leq 0$, and $f(q)= 0$, otherwise, satisfies all the above equations with $\phi=0$. Such a deformation, representing two flat regions connected by a plastic fold with no stress and moment in $\Omega$, is a solution to our problem in the absence of any external forcing in the bulk or at the boundary. We now obtain the solutions when moment and transverse force are prescribed at one edge of the domain. We consider a rectangular domain $\Omega$ to be defined by $a_0\leq q\leq a_1$ such that $a_0\leq 0 \leq a_1$. All the boundaries of $\Omega$, except at $x_c=a_1$, are assumed to be free of forces and moments. At $x_c=a_1$, we take $\phi=0$, ${\partial \phi}/{\partial q}=0$, $D{d^2w}/{d{q}^2}=b_0$, and $D{d^3w}/{d{q}^3}=b_1$. The first two conditions impose that there are no in plane forces applied at the edge. The constants $b_0$ and $b_1$ represent the applied moment and the applied transverse force at the edge. The deformation $f(q)=0$, if $q \leq 0$, and $f(q) = -\gamma_0 q +k_1 {q}^2  + k_2 {q}^3$, otherwise, satisfies all the governing equations and the boundary conditions with $\phi=0$, $k_1=b_0/2D - b_1a_1/2D$, and $k_2=b_1/6D$. 

As the second problem, we consider a circular interface $S$ given by $|\boldsymbol{x}|=r_0$, with normal $\boldsymbol{\nu}=\boldsymbol{e}_r$ and curvature $k={1}/{r_0}$, see Figure~\ref{cirfold}. The fold is taken as $\boldsymbol{\gamma}=\gamma_0 \boldsymbol{e}_r \otimes\boldsymbol{e}_r$, where $\gamma_0$ is constant. We consider a transverse displacement, compatible with the fold, as $w =  \gamma_0 r_0$, if $r \leq r_0$, and $w= \gamma_0 r$, if $r > r_0$, i.e., the fold deforms the sheet into a planar disc-like region (of radius $r_0$) and a cone for $r > r_0$, see Figure~\ref{cirfold}. We assume a plastic stretching strain of the form $\boldsymbol{e}^p=\boldsymbol{0}$, if $r \leq r_0$, and $\boldsymbol{e}^p = {\gamma_0}^2 \boldsymbol{e}_r\otimes \boldsymbol{e}_r$, if  $r > r_0$.
The assumed form of $w$ and $\boldsymbol{e}^p$ satisfy the conditions~\eqref{foldcomp} identically. Note that we need a non trivial plastic stretching strain to satisfy the interfacial compatibility condition \eqref{FoldCylindricalInterfacialCompatibility1} for the given deformation under the assumption of elastic inextensibility. On the other hand, the equilibrium conditions \eqref{FoldBulkEquilibrium} and \eqref{FoldInterfacialEquilibrium1} are satisfied for $\phi=-\gamma_0 \ln r$. The interfacial equilibrium condition \eqref{FoldInterfacialEquilibrium2} however can not be satisfied in the present form. In order to satisfy \eqref{FoldInterfacialEquilibrium2} for the given $w$ we would require a distribution of transverse force couple field over $S$ in the form $f_2=-D\gamma_0/r_0$. Therefore a sheet of paper can be folded about a circular curve, with a flat disc like region in the interior and a perfect cone shape on the other side, if we superpose a distribution of  transverse force couples along the fold.

\subsection{Fold terminating at an internal point} \label{ridge}

\begin{figure}[t!]
\begin{subfigure}{.49\textwidth}
  \centering
\includegraphics[scale=0.55]{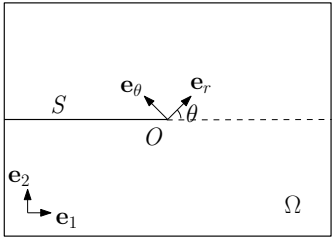}
\vspace{20pt}
\subcaption{The plate domain.} \label{singfold1}
\end{subfigure}%
%\hspace{35pt}
\begin{subfigure}{.49\textwidth}
  \centering
\includegraphics[scale=0.3]{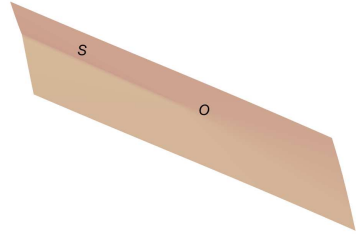}
\vspace{-20pt}
\subcaption{The transverse displacement.} \label{singfold2}
\end{subfigure}%
\caption{The plate domain showing a straight fold $S$ terminating at $O$ and the corresponding transverse displacement.}
\label{singfold}
\end{figure}

We consider a linear fold $S$ terminating in the interior of the domain $\Omega$ at $O$, see Figure~\ref{singfold1}. In terms of the polar coordinates $r \geq 0$, $-\pi < \theta \leq \pi$, with origin at $O$, the fold lies on the $\theta = \pi$ line. The constant tangent $\boldsymbol{t}$ to $S$ is along $\boldsymbol{e}_1$ while the constant normal $\boldsymbol{\nu}$ to $S$ is along $\boldsymbol{e}_2$. We assume the stretching plastic strain to vanish identically over $\Omega$ and the bending plastic strain to concentrate on $S$, i.e., $\boldsymbol{E}^p=\boldsymbol{0}$ and $\boldsymbol{\Lambda}^p \in \mathcal{C}(\Omega,\Sym)$, such that $\boldsymbol{\Lambda}^p(\boldsymbol{\psi})=\gamma_0 \int_S \langle \boldsymbol{e}_2\otimes \boldsymbol{e}_2,\boldsymbol{\psi} \rangle \dl$ for all $\boldsymbol{\psi} \in \mathcal{D}(\Omega,\Lin)$, where $\gamma_0$ is constant. Such a plastic strain distribution yields a point supported incompatibility, $\Curl \boldsymbol{\Lambda}^p=-\gamma_0 \boldsymbol{e}_2 \delta_O$. In other words, there is no interfacial incompatibility and the incompatibility concentrates at the end point of the fold in the interior of the domain. In comparison, the conical singularity (in Section~\ref{conical}) allows for an incompatibility through a non-trivial Gaussian curvature at the defect point while the circular fold (in Section~\ref{folds}) has an incompatibility concentrating on the fold.

For the considered form of $\boldsymbol{\Lambda}^p$, and recalling that there is no concentration in elastic bending strain, we obtain $\llbracket \nabla w \rrbracket=-\gamma_0 \boldsymbol{\nu}$. Consequently, we use elastic inextensibility (i.e., $\boldsymbol{E}^e = \boldsymbol{0}$ in $\Omega$) and $k=0$ to simplify the first von K{\'a}rm{\'a}n equation \eqref{vk1loc} as
\begin{subequations}
\begin{align}
\label{RidgeCompatibilityBulk}
[w,w]=0 ~\text{in }~\Omega-\{S \cup \{O\} \},
\\
\label{RidgeCompatiiblityInterface}
\langle \{ \nabla \nabla w\},\boldsymbol{e}_1 \otimes\ \boldsymbol{e}_1 \rangle \gamma_0 =0 ~\text{on }~S-\{O\}, ~\text{and}
\\
\label{RidgeCompatibiltiySingular}
\nabla W \otimes \nabla W(\mathbb{A}\nabla\nabla v^\alpha)=0 ~\text{at}~O,
\end{align}
\end{subequations}
for all multi indices $\alpha \in \mathbb{N}^2$ such that $|\alpha|\leq\deg(\nabla W \otimes \nabla W)+2$. On the other hand, the second von K{\'a}rm{\'a}n equation~\eqref{VonKArmanEquilibriumStrongForms}, with a  vanishing body force field, simplifies to
\begin{subequations}
\label{RidgeEquilibrium}
\begin{align}
\label{RidgeEquilibriumBulk}
D\Delta^2 w - [\phi,w]=0 ~ \text{in}~\Omega-\{S\cup\{O\}\},
\\
\label{RidgeEquilibriumInterface1}
\begin{aligned}
D\left\langle \llbracket \div \nabla\nabla w \rrbracket, \boldsymbol{e}_2 \right\rangle + \gamma_0 \left\langle \nabla\nabla \phi, \boldsymbol{e}_1 \otimes \boldsymbol{e}_1 \right\rangle  = 0  ~\text{on} ~S-\{O\}, 
\end{aligned}
\\
\label{RidgeEquilibriumInterface2}
\llbracket\Delta w\rrbracket =  0  ~\text{on} ~S-\{O\},~\text{and}
\\
\label{RidgeEquilibriumSingular}
D W(\Delta^2 v^\alpha)+\nabla \Phi \otimes \nabla W(\mathbb{A}\nabla\nabla v^\alpha)=-D\left( (1-\nu)\boldsymbol{\Lambda}^p(\nabla\nabla v^\alpha) +\nu  (\tr(\boldsymbol{\Lambda}^p)(\Delta v^\alpha))\right)~\text{at}~O,
\end{align}
\end{subequations}
for all multi indices $\alpha \in \mathbb{N}^2$ such that $|\alpha| \leq q$ where $q$ is the maximum of $\deg(\nabla \Phi \otimes \nabla W)+2$, $\deg(W)+4$, and $\deg(\boldsymbol{\Lambda}^p)+2$.

We consider a transverse deformation of the form $w=rg(\theta)$, where $g$ is continuous but piecewise smooth across $S$.  The condition $\llbracket \nabla w \rrbracket=-\gamma_0 \boldsymbol{\nu}$ then reduces to 
\begin{equation}
\label{RidgeJumpNormal}
\frac{dg}{d\theta}(\pi)-\frac{dg}{d\theta}(-\pi)=\gamma_0.
\end{equation}
The considered $w$ satisfies \eqref{RidgeCompatibilityBulk} identically.  On substituting it in \eqref{RidgeCompatibiltiySingular} we obtain the condition
\begin{equation}
\label{RidgeLoopIntegralCompatibiltiyG}
\int_{-\pi}^{\pi} g\left(g+\frac{d^2g}{d\theta^2}\right) d\theta + g(\pi) \left( \frac{dg}{d\theta}(\pi)-\frac{dg}{d\theta}(-\pi) \right)=0.
\end{equation} 
On using the stress function $\phi=D\lambda\ln r$, \eqref{RidgeEquilibriumBulk} reduce to
\begin{equation}
\label{RidgeEquilibriumEquG}
\frac{d^4g}{d\theta^4}+(\lambda+2)\frac{d^2g}{d\theta^2}+(\lambda+1)g=0.
\end{equation}
This equation has a solution $g(\theta)=a \cos \mu\theta$, where $\mu^2 = \lambda + 1$. It should be noted that here, unlike the solution in Section~\ref{conical}, $\mu$ is not necessarily an integer. The constant coefficient $a$ is obtained by substituting this solution in \eqref{RidgeJumpNormal} as $a = {\gamma_0}/({2\mu \sin(\pi \mu )})$.  To calculate $\mu$, we substitute the solution for $g$ in  \eqref{RidgeLoopIntegralCompatibiltiyG} to obtain 
\begin{equation}
\label{RidgeMuEquation}
(1+\mu^2)\sin(2\pi \mu)+2\pi (1-\mu^2)\mu =0.
\end{equation}
The nonlinear equation can be solved (numerically) to obtain $\mu = 0.92$ (the other roots are imaginary). Altogether, we have a transverse displacement of the form~\cite{lechenault2015generic}
\begin{equation}
w=\frac{\gamma_0 r}{2\mu \sin(\pi \mu )}\cos \mu\theta,
\end{equation}
where $\mu = 0.92$, see Figure~\ref{singfold2}. The considered solutions for $w$ and $\phi$ identically satisfy \eqref{RidgeCompatiiblityInterface}, \eqref{RidgeEquilibriumInterface1} and \eqref{RidgeEquilibriumInterface2}. However Equation \eqref{RidgeEquilibriumSingular}, which enforces equilibrium of transverse forces at $O$, is not satisfied. Whereas the left hand side terms reduce to a Dirac, the right hand side terms reduce to a gradient of the Dirac. The latter is proportional to $\nu$ while the former is not. This inconsistency of the considered solution in its failure to satisfy the equilibrium equation at $O$ indicates that the assumption of elastic inextensibility needs to be perhaps relaxed in the neighborhood of the singular point.

\subsection{Folding into a tetrahedron} \label{tetra}

\begin{figure}[t!]
\begin{subfigure}{.49\textwidth}
  \centering
\includegraphics[scale=0.40]{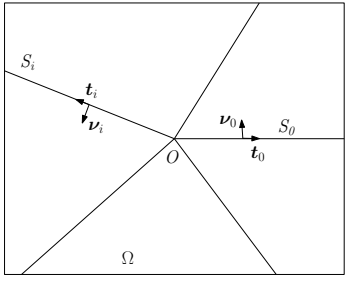}
\vspace{20pt}
\subcaption{The plate domain.} \label{tetra1}
\end{subfigure}%
%\hspace{35pt}
\begin{subfigure}{.49\textwidth}
  \centering
\includegraphics[scale=0.3]{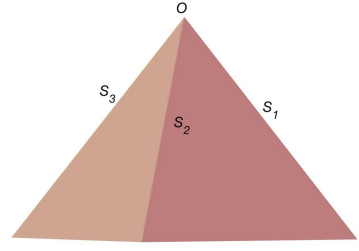}
%\vspace{-20pt}
\subcaption{The transverse displacement.} \label{tetra2}
\end{subfigure}%
\caption{The plate domain showing several straight folds $S$ terminating at $O$ on one end and at $\partial \Omega$ on the other. The  transverse displacement for a given prescription of folds in terms of plastic bending strain field.}
\end{figure}

We consider a finite number ($n$) of linear folds each terminating at a point $O$ in the interior of the domain, see Figure~\ref{tetra1}. Let $S=\cup_{i=1}^n S_i$, where $S_i$ represents the $i$-th fold. Each $S_i$ is a straight line with one end point at $\partial \Omega$ and the other end point at $O$. The multitude of folds terminating in the interior of the domain are modelled through a plastic bending strain concentrated on $S$ such that $\boldsymbol{\Lambda}^p(\boldsymbol{\psi})=\sum_{i=0}^{n-1} \gamma_i \int_{S_i}\langle \boldsymbol{\nu}_i \otimes \boldsymbol{\nu}_i,\boldsymbol{\psi} \rangle \dl$ where $\gamma_i$ and $\boldsymbol{\nu}_i$ represent the (constant) strength and the (constant) normal associated with the $i$-th fold. We assume that the incompatibility $N_1$ is identically zero and the plastic bending strain is compatible, i.e., $\Curl \boldsymbol{\Lambda}^p=\boldsymbol{0}$, which is equivalent to $\sum_{i=0}^{n-1} \gamma_i \boldsymbol{\nu}_i =\boldsymbol{0}$. Due to a compatible plastic strain overall, we look for a solution such that there are no elastic strain fields necessary to accommodate the prescribed plastic strains. There are then no stretching strains and the total bending strain is identical to the plastic bending strain. The transverse displacement field $W$ satisfies $\boldsymbol{\Lambda}^p = \nabla \nabla W$. All the governing equations, including the compatibility conditions and equilibrium conditions, are trivially satisfied. In the following lemma, we calculate the Gaussian curvature that appears due to such a bending strain distribution. It is point supported at $O$ with only a Dirac, similar to the case of isolated disclinations. The source of Gaussian curvature presently, however, is in the plastic folds imposed on the domain.

\begin{lemma}
Given $\boldsymbol{\Lambda}^p \in \mathcal{C}(\Omega,\Sym)$ such that $\boldsymbol{\Lambda}^p(\boldsymbol{\psi})=\sum_{i=0}^{n-1} \gamma_i \int_{S_i}\langle \boldsymbol{\nu}_i \otimes \boldsymbol{\nu}_i,\boldsymbol{\psi} \rangle \dl$, where $S_i$ represents a straight line satisfying $\partial S_i - \partial \Omega=\{O\}$ and $\boldsymbol{\nu}_i$ is the normal to $S_i,$ and $\Curl \boldsymbol{\Lambda}^p=\boldsymbol{0}$, we obtain
\begin{subequations}
\begin{align}
\label{TetraCurl}
\sum_{i=0}^{n-1} \gamma_i \boldsymbol{\nu}_i =\boldsymbol{0}~\text{and}
\\
\label{TetraGaussian}
\Det (\boldsymbol{\Lambda}^p)=-\frac{1}{2}\sum_{j=1}^{n-1}\sum_{i=0}^{j-1} \gamma_i \gamma_j \langle \boldsymbol{\nu}_i,\boldsymbol{t}_j \rangle \delta_O,
\end{align}
\end{subequations} 
where $\boldsymbol{t}_j$ is the tangent to $S_j.$
\begin{proof}
The relation \eqref{TetraCurl} follows immediately from \cite[Id. 2.2]{pandey2020topological}. Let $\Omega_i$ be the angular sector bounded by $S_{i}$ and $S_{i+1}.$ Let $W \in \mathcal{B}(\Omega)$ such that $\nabla W\in \mathcal{B}(\Omega,\mathbb{R}^2)$ with bulk density $\nabla w = -\sum_{j=0}^{i} \gamma_j \boldsymbol{\nu}_j$ in $\Omega_i.$ We use \cite[Id. 2.2]{pandey2020topological} to obtain
\begin{equation}
\Curl\Curl (\nabla W\otimes \nabla W)= \sum_{j=1}^{n-1}\sum_{i=0}^{j-1} \gamma_i \gamma_j \langle \boldsymbol{\nu}_i,\boldsymbol{t}_j \rangle \delta_O,
\end{equation}
which establishes the desired result.
\end{proof}
\end{lemma}
The transverse displacement $w$ is illustrated in Figure~\ref{tetra2}.

\section{Conclusion}
In this work we have extended the classical von K{\'a}rm{\'a}n plate equations, given in terms of smooth fields, to a distributional form while allowing the fields to be piecewise smooth across curves, to concentrate on the curves, and to be singular at points in the domain. In writing the distributional form, the central challenge was to unambiguously define the notions of distributional determinant, distributional inner product, and distributional Monge-Amp{\`e}re bracket. The sources of inhomogeneities were taken in the form of plastic strains, incompatibility fields (which in turn are related to the defect densities), and body force fields, all allowed to be singular over points and curves in the plate domain. The distributional forms of the von K{\'a}rm{\'a}n plate equations were localized to obtain pointwise local forms of the equations in the bulk, away from the singular points and curves, on the singular curves, and at the singular points. These local equations are the main contribution of this paper. The applicability of the developed framework was illustrated through several examples associated with conical deformations (arising due to defects and constrained deformations), folds, and folds terminating inside the plate domain at a singular point. Our work provides a complete and rigorous framework where a large variety of singular problems associated with von K{\'a}rm{\'a}n plates can be discussed and systematically analyzed. 

\section*{Ackowledgement}
AG acknowledges the financial support from SERB (DST) Grant No. CRG/2018/002873 titled ``Micromechanics of Defects in Thin Elastic Structures".

\bibliographystyle{plain}
\bibliography{ref}

\end{document}